\documentclass[preprint,aps,showpacs]{revtex4}
\usepackage{mathrsfs}
\usepackage{amsmath}
\usepackage{amssymb}
\usepackage{epsfig}
\usepackage{graphicx}
\usepackage{booktabs}
\usepackage{multirow}
\usepackage{color}
\begin{document}
\renewcommand{\arraystretch}{0.5}
\newcommand{\beq}{\begin{eqnarray}}
\newcommand{\eeq}{\end{eqnarray}}
\newcommand{\non}{\nonumber\\ }

\newcommand{\acp}{ {\cal A}_{CP} }
\newcommand{\psl}{ p \hspace{-1.8truemm}/ }
\newcommand{\nsl}{ n \hspace{-2.2truemm}/ }
\newcommand{\vsl}{ v \hspace{-2.2truemm}/ }
\newcommand{\epsl}{\epsilon \hspace{-1.8truemm}/\,  }

\def \cpl{ Chin. Phys. Lett.  }
\def \ctp{ Commun. Theor. Phys.  }
\def \epjc{ Eur. Phys. J. C }
\def \jpg{  J. Phys. G }
\def \npb{  Nucl. Phys. B }
\def \plb{  Phys. Lett. B }
\def \prd{  Phys. Rev. D }
\def \prl{  Phys. Rev. Lett.  }
\def \zpc{  Z. Phys. C }
\def \jhep{ J. High Energy Phys.  }

\title{ S-wave ground state charmonium decays of $B_c$ mesons in the
perturbative QCD approach}
\author{Zhou Rui$^a$}\email{zhourui@heuu.edu.cn}
\author{Zhi-Tian Zou$^b$}
\affiliation{a. School of Science, Hebei United University, Tangshan 063009,  People¡¯s Republic of China\\
b. Department of Physics, Yantai University, Yantai 264005,   China}

\date{\today}
\begin{abstract}
We make a systematic investigation on the two-body nonleptonic decays $B_c\rightarrow J/\Psi(\eta_c),M$
by employing the perturbative QCD approach based on $k_T$ factorization, where $M$ is a light
 pseudoscalar  or  vector   or a heavy charmed meson. We predict the branching ratios and direct
CP asymmetries of these $B_c$
decays and also the transverse polarization fractions of $B_c\rightarrow J/\Psi V,J/\psi D^{*}_{(s)}$ decays.
It is found that  these decays have a large branching ratios of the order of $10^{-4} - 10^{-2}$ and  could be  measured by the future LHC-b experiment. Our predictions for the  ratios of branching fractions $\frac{\mathcal {BR}(B_c^+\rightarrow J/\Psi D_s^+)}
{\mathcal {BR}(B_c^+\rightarrow J/\Psi \pi^+)}$ ,$\frac{\mathcal {BR}(B_c^+\rightarrow J/\Psi D_s^{*+})}
{\mathcal {BR}(B_c^+\rightarrow J/\Psi D_s^+)}$ and $\frac{\mathcal {BR}(B_c^+\rightarrow J/\Psi K^+)}
{\mathcal {BR}(B_c^+\rightarrow J/\Psi \pi^+)}$ are in good agreement with the data.
A large  transverse polarization fraction which can reach $48\%$ is predicted in
 $B_c^+\rightarrow J/\Psi D_s^{*+}$ decay, which is consistent with the data.
 We find a possible  direct CP violation in $B_c\rightarrow  J/\psi D^{*}$ decays,
  which are helpful to test the CP violating effects in $B_c$ decays.

\end{abstract}



\maketitle

\section{Introduction}

Since the first discovery of the $B_c$ meson by the CDF collaboration at Tevatron in 1998 through the semileptonic
modes $B_c\rightarrow J/\psi (\mu^+\mu^-)l^+X(l=e,\mu)$\cite{prl812432}, it has aroused a great deal of interest
in studying  $B_c$ physics experimentally. Subsequent measurements of its mass  and lifttime in different detectors via
the two processes $B_c^+\rightarrow J/\psi l^+\nu_l$  \cite{prl012002,prl092001} and
$B_c^+\rightarrow J/\psi \pi^+$ \cite{prl182002,prl012001} have opened new windows for the analysis of the  dynamics involved
in the $B_c$ decays. At the current level accuracy, around $5\times 10^{10}$ $B_c$ events are expected to be produced each year
\cite{jmpa5117}. Up to now, the LHCb  collaboration has measured the $B_c$ mass with $6273 \pm 1.3 (stat) \pm 1.6 (syst) \text{MeV}/\text{c}^2$
\cite{prl232001} and some new channels, such as $B_c^+\rightarrow J/\psi \pi^+ \pi^- \pi^+$ \cite{prl281802}, $B_c^+\rightarrow J/\psi K^+$ \cite{jhep075},
 $B_c^+\rightarrow \psi(2S) \pi^+$ \cite{prd071103}, $B_c^+\rightarrow J/\psi D_s^{(*)+}$ \cite{prd112012}, $B_c^+\rightarrow J/\psi K^+K^-\pi^+$ \cite{jhep094},$B_c^+\rightarrow B_s^0\pi^+$ \cite{prl181801}, and  $B_c^+\rightarrow J/\psi 3\pi^+2\pi^-$ \cite{14040287} for the first time. We can see all of the  observed processes involving the $J/\psi$ final state, due to the narrow peak of $J/\psi$  and
 the high purity of $J/\psi\rightarrow l^+l^-$, the decay modes containing the  signal of $J/\psi$ meson are among the most easily reconstructible $B_c$ decay
 modes.  One should expect that, in the following years, more and more charmonium decay modes of $B_c$ meson will be measured with good precision in the LHCb experiments.

Compared with the $B_{u,d,s}$ mesons, the $B_c$ meson is of special interest. Being the ground  state of two  heavy quarks
of different flavors ($\bar{b}$ and $c$ ), $B_c$ decays via weak interaction  only, while the strong and
electromagnetic annihilation processes  are forbidden. Since both of the two quarks are heavy, each of them
can decay with the other as a spectator, the $B_c$ meson have much shorter lifetime than other
b-flavored mesons \cite{prd011101}, pointing to the important role of the $c$ quark in $B_c$ decays.
 It has rich decay channels, and provides a very good place to study
nonleptonic weak decays of heavy mesons, to test the standard model
and to search for any new physics signals \cite{iiba}.

Theoretically, many hadronic $B_c$ decay modes have been studied by various theoretical approaches.
The perturbative QCD approach (pQCD)\cite{prl744388}  is one of
the recently developed theoretical tools based on QCD to
deal with the nonleptonic $B$ decays. Utilizing the $k_T$ factorization instead of collinear factorization, this approach is
free of end-point singularity. Thus the Feynman diagrams,
including factorizable, nonfactorizable, and annihilation
type, are all calculable. Up to now, the pure annihilation type of charmless $B_c\rightarrow PP,PV,VV,AV,AA,AP,SP,SV$
decays \cite{prd014022,prd074017,prd074012,prd054029,jpg035009,prd074033,14010151} and the
charm decays of $B_c\rightarrow D^{(*)}_{(s)}(P,V,T,D^{(*)}_{(s)})$ \cite{epjc45711,epjc63435,prd074008,prd074019,prd074027}
have been studied systematically in the pQCD approach, where the term
$P,V,A,S,T$ refers to the pesudoscalar, vector, axial-vector, scalar and tensor charmless mesons, respectively.

In the present paper, we extend our pQCD analysis
to the S-wave ground state charmonium decays of the $B_c$ meson. The $B_c\rightarrow J/\psi(\eta_c)\pi$ \cite{epjc60107}, $B_c\rightarrow J/\psi K$ \cite{prd037501}
decays have been studied in pQCD, compared to which the new ingredients of this
paper are: (1) we  updated the Cabibbo-Kobayashi-Maskawa (CKM)
matrix elements and  some input hadronic parameters according to the Particle Data Group 2012 \cite{pdg};
(2)we have included the intrinsic $b$
( the conjugate space coordinate of the parton transverse momentum $k_T$)
dependence for the $B_c$ meson wave function,
because it is observed that the intrinsic $b$  dependence in the heavy meson wave functions is  important \cite{prd2480};
(3) not only the $B_c\rightarrow J/\psi(\eta_c),\pi(K)$ decays, but $B_c\rightarrow (J/\psi,\eta_c)(\pi,K,K^*,\rho,D^{(*)}_{(s)})$
are investigated. In addition, a comprehensive study of  these processes,   which have been studied  in the QCD coupling \cite{prd034008},
the relativistic quark model  \cite{prd094020},
the covariant light-front quark model \cite{epjc51841} and so on, is still lacking
in pQCD. Our aim is to fill in this gap and provide a ready
reference to the existing and forthcoming experiments
to compare their data with the predictions in the pQCD
approach. It will be shown that the obtained ratios of the branching ratios and  polarization
fractions are all in consistency with the existing data.

In the $B_c$ rest frame, since both of the constituents (c, $\bar{b}$) are heavy, they are almost at rest relative to each other. The $B_c$ meson can be approximated as a nonrelativistic quarkonium system \cite{jhep04061,10126007}. In this sense  the charm quark mass, which is considerably larger than the QCD scale, provides an intrinsic physical infrared regulator. The
dynamics at this scale is still calculable perturbatively \cite{jhep04061}.
In the  pQCD framework, since the  spectator charm quark  is
almost at rest, a hard gluon is needed to transfer
energy to make it a collinear quark into the final state meson.
 Meanwhile, the heavy charm mass will bring another expansion series of $m_c/m_{B_c}\sim 0.2$. In fact, the factorization theorem is applicable to the
$B_c$ system similar to the situation of the  $B$  meson \cite{prd074033} in the leading order of this expansion.
For the decays with a  heavy charmonium and a light meson in the final states,
since the emitted meson is a light meson,
the factorization could be proved  in the soft-collinear effective theory  to all orders of the strong coupling constant in the
heavy quark limit \cite{epjc51841,scet1}.
For the decays with a  heavy charmonium and a charm meson in the final states, both the charmonium and  charm meson can
emit from the weak vertex,
which is similar to the double charm decays of the  $B_c$ meson \cite{prd074019}. The proof of factorization here is thus trivial.
In fact, this type of  process in $B$ meson decays has been studied in the pQCD approach successfully \cite{liying}.

Our paper is organized as follows: We review the pQCD factorization
approach and then   perform the perturbative calculations for these
considered decay channels   in Sec.\ref{sec:f-work}.  The numerical
results and discussions on the observables are given in
Sec.\ref{sec:result}. The final section is devoted to our
conclusions. Some details of related functions and the decay amplitudes
are given in the Appendix.

\section{  Framework and wave function}\label{sec:f-work}
At the quark level, the considered processes are characterized by the $\bar{b}\rightarrow \bar{c} q\bar{q}'$ transition,
with $q=u,c$ and $\bar{q}'=\bar{d},\bar{s}$. In the rest frame of  the $B_c$ meson, the spectator $c$ quark is almost at rest due to the heavy mass.
Therefore, a hard gluon is then needed to transform the $c$ quark into a collinear object in the
final charmonium or charmed meson.  This makes the perturbative calculations into a six-quark interaction.
These perturbative calculations meet end-point singularity
in dealing with the meson distribution amplitudes at the end point. We take back the parton
transverse momentum $k_T$ to regulate this divergence.
In the pQCD approach, the decay amplitude can be written as the following factorizing formula \cite{prd5577},
\begin{eqnarray}\label{eq:factorization}
C(t)\otimes H(x,t)\otimes
\Phi(x)\otimes\exp[-s(P,b)-2\int^t_{1/b}\frac{d\mu}{\mu}\gamma_q(\alpha_s(\mu))],
\end{eqnarray}
where $C(t)$ is Wilson coefficient of the four-quark operator with
the QCD radiative corrections. $t$ is chosen as the largest energy scale in
the hard part, in order to lower  the largest logarithm.
The term $\exp[-s(P, b)]$ \cite{npb193381}, the so-called Sudakov factor, results from summing up
double logarithms caused by collinear divergence and soft divergence, with P denoting the dominant
light-cone component of meson momentum. $\gamma_q=-\alpha_s/\pi$ is the quark anomalous dimension.
The hard part $H(x,t)$  can be perturbatively calculated including all possible Feynman diagrams
 without end-point singularity, such as  factorizable, nonfactorizable and
annihilation-type diagrams.  The wave function $\Phi(x)$, which describes hadronization of the
quark and antiquark to the meson, is not calculable and treated as nonperturbative inputs.

The meson wave function absorbs nonperturbative dynamics of the process, which is process independent.
Using the wave functions determined from other well-measured processes, one can make quantitative predictions here.
Similar to the situation of $B$ meson, for  $B_c$ meson, one of the
dominant Lorentz structure is considered in the
numerical calculations,  while the contribution induced
by the other Lorentz structures is  negligible \cite{epjc28515}.
In the nonrelativistic limit, we use the same distribution amplitude for $B_c$
meson as those used in Refs. \cite{prd074008,prd074019,prd074027}
\begin{eqnarray}
\Phi_{B_c}(x)=\frac{if_B}{4N_c}[(\rlap{/}{P}+M_{B_c})\gamma_5\delta(x-\frac{m_c}{M_{B_c}})]\exp(-\frac{\omega_B^2 b^2}{2}),
\end{eqnarray}
in which the last exponent term  represents  the $k_T$ dependence. The shape parameter $\omega_B=0.60\pm0.05$ GeV has been adopted
in  our previous analyses of the double charm decays of $B_c$ meson \cite{prd074019}.

The two-particle light-cone distribution
amplitudes of the $D_{(s)}/D_{(s)}^*$ meson can be written as \cite{prd67054028}
\begin{eqnarray}\label{eq:dwave}
\langle D_{(s)}(P_2)|q_{\alpha}(z)\bar{c}_{\beta}(0)|0\rangle &=&
\frac{i}{\sqrt{2N_c}}\int^1_0dx e^{ixP_2\cdot
z}[\gamma_5(\rlap{/}{P}_2+m_{D_{(s)}})\phi_{D_{(s)}}(x,b)]_{\alpha\beta},
\nonumber\\
\langle D_{(s)}^*(P_2)|q_{\alpha}(z)\bar{c}_{\beta}(0)|0\rangle
&=&-\frac{1}{\sqrt{2N_c}}\int^1_0dx e^{ixP_2\cdot
z}[\rlap{/}{\epsilon}_L(\rlap{/}{P}_2+m_{D_{(s)}^*})\phi^L_{D_{(s)}^*}(x,b)\nonumber\\&&
+\rlap{/}{\epsilon}_T(\rlap{/}{P}_2+m_{D_{(s)}^*})\phi^T_{D_{(s)}^*}(x,b)]_{\alpha\beta},
\end{eqnarray}
with the  normalization conditions:
\begin{eqnarray}
\int^1_0dx \phi_{D_{(s)}}(x,0)=\frac{f_{D_{(s)}}}{2\sqrt{2N_c}},\quad  \int^1_0dx
\phi^{L}_{D_{(s)}^*}(x,0)= \int^1_0dx
\phi^{T}_{D_{(s)}^*}(x,0)=\frac{f_{D_{(s)}^*}}{2\sqrt{2N_c}}.
\end{eqnarray}
Here we  use $f_{D_{(s)}^*}=f^T_{D_{(s)}^*}$ in the calculation.
The value of  $f_{D^*_{(s)}}$ is   determined by
the following relations derived from HQET \cite{hqet}:
\begin{eqnarray}\label{eq:Ddecayc}
f_{D^*_{(s)}}=\sqrt{\frac{m_{D_{(s)}}}{m_{D_{(s)}^*}}}f_{D_{(s)}}.
\end{eqnarray}
The distribution amplitude  $\phi^{(L,T)}_{D_{(s)}^{(*)}}$ is taken as \cite{09101424}
\begin{eqnarray}
\phi^{(L,T)}_{D_{(s)}^{(*)}}=\frac{3}{\sqrt{2N_c}}f_{D^{(*)}_{(s)}}x(1-x)[1+a_{D^{(*)}_{(s)}}(1-2x)]\exp(-\frac{\omega^2_{D_{(s)}}b^2}{2}).
\end{eqnarray}
 According to Ref. \cite{prd034006}, we use $a_D=0.5 \pm 0.1, \omega_{D}= 0.1 \text{GeV}$ for the $D/D^*$ meson and $a_D=0.4 \pm 0.1, \omega_{D_s}= 0.2 \text{GeV}$ for the  $D_s/D_s^*$ meson.

For the  $J/\psi(\eta_c)$ meson, in terms of the notation in Ref.\cite{prd114008}, we
decompose the nonlocal matrix elements for the longitudinally and transversely polarized $J/\psi$ mesons and $\eta_c$
into
\begin{eqnarray}
\langle J/\psi (P, \epsilon^L)|\bar{c}(z)_{\alpha}c(0)_{\beta}|0\rangle &=& \frac{1}{\sqrt{2N_c}}\int_0^1 dxe^{ixP\cdot z}
[m_{J/\psi}\rlap{/}{\epsilon^L}_{\alpha\beta}\psi^L(x,b)+(\rlap{/}{\epsilon^L}\rlap{/}{P})_{\alpha\beta}\psi^t(x,b)], \nonumber\\
\langle J/\psi (P, \epsilon^T)|\bar{c}(z)_{\alpha}c(0)_{\beta}|0\rangle &=& \frac{1}{\sqrt{2N_c}}\int_0^1 dxe^{ixP\cdot z}
[m_{J/\psi}\rlap{/}{\epsilon^T}_{\alpha\beta}\psi^V(x,b)+(\rlap{/}{\epsilon^T}\rlap{/}{P})_{\alpha\beta}\psi^T(x,b)],\nonumber\\
\langle \eta_c (P)|\bar{c}(z)_{\alpha}c(0)_{\beta}|0\rangle &=& -\frac{i}{\sqrt{2N_c}}\int_0^1 dxe^{ixP\cdot z}
[(\gamma_5\rlap{/}{P})_{\alpha\beta}\psi^v(x,b)+m_{\eta_c}(\gamma_5)_{\alpha\beta}\psi^s(x,b)],\nonumber\\
\end{eqnarray}
respectively.  $\psi^L$, $\psi^T$ and $\psi^v$ denote for the twist-2 distribution amplitudes, while $\psi^t$, $\psi^V$ and $\psi^s$ for the
twist-3 distribution amplitudes. $x$ represents the momentum fraction of the charm quark inside the
charmonium.  In order to include the intrinsic $b$ dependence for the $J/\psi(\eta_c)$ meson wave function,
 we adopt the same model as \cite{epjc60107}.
 For the  wave functions of light vector and pseudoscalar  mesons,  the same form and parameters are adopted as \cite{prd074008} and one is referred to the original
 literature \cite{wave}.
\subsection{ $B_c\rightarrow (J/\psi, \eta_c)(P, V)$ decays}
\begin{figure}[!htbh]
\begin{center}
\vspace{-2cm} \centerline{\epsfxsize=12 cm \epsffile{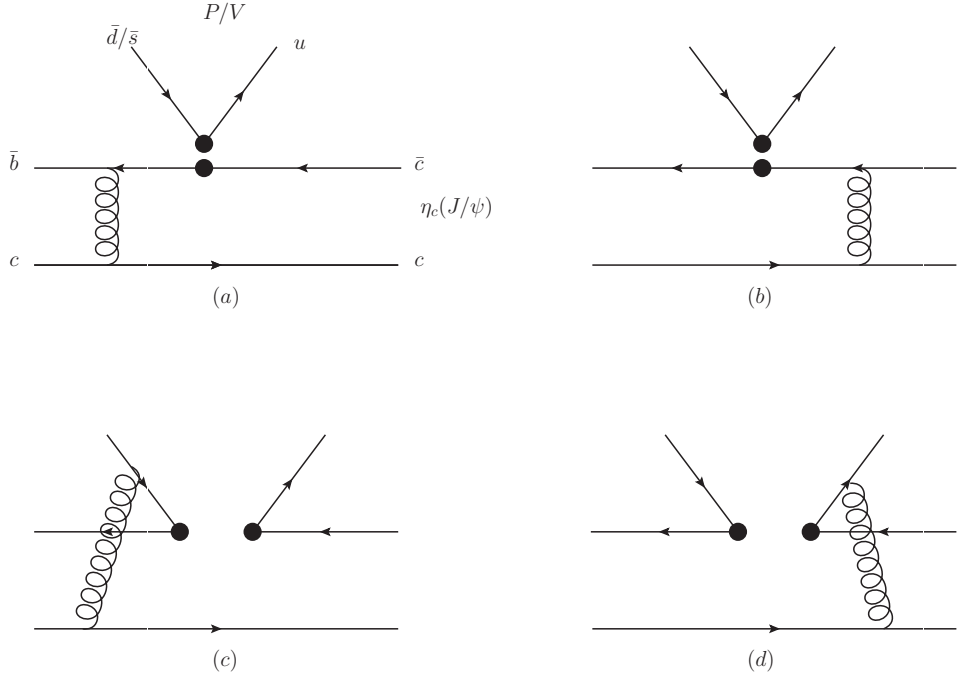}}
\vspace{-6.6cm} \caption{Feynman diagrams for  $B_c\rightarrow (J/\psi, \eta_c)(P, V)$ decays.}
 \label{fig:jpv}
 \end{center}
\end{figure}
The effective Hamiltonian for these modes can be written as
\begin{eqnarray}
\mathcal{H}_{eff}=\frac{G_F}{\sqrt{2}}V^*_{cb}V_{ud(s)}(C_1(\mu)O_1(\mu)+C_2(\mu)O_2(\mu)),
\end{eqnarray}
with $V^*_{cb}$ and $V_{ud(s)}$  the Cabibbo-Kobayashi-Maskawa (CKM) matrix elements, $C_{1,2}(\mu)$
 the perturbatively calculable Wilson coefficients, and $O_{1,2}(\mu)$
 the effective four-quark operators; their expressions are
\begin{eqnarray}
O_{1}(\mu)&=&\bar{b}_{\alpha}\gamma^{\mu}(1-\gamma_5)c_{\beta} \otimes \bar{u_{\beta}}\gamma_{\mu}(1-\gamma_5)q'_{\alpha},\nonumber\\
O_{2}(\mu)&=&\bar{b}_{\alpha}\gamma^{\mu}(1-\gamma_5)c_{\alpha}\otimes \bar{u}_{\beta}\gamma_{\mu}(1-\gamma_5)q'_{\beta},
\end{eqnarray}
where $\alpha,\beta$ are  color indices and the summation convention over repeated indices is understood.
Since the four quarks in the operators are different from each other, there is no penguin contribution, and thus there is
no $CP$ violation. With the effective Hamiltonian given above, the Feynman diagrams corresponding to the concerned process are
drawn in Fig.\ref{fig:jpv}  where the first two are of factorizable topology  contributing to the form factor of $B_c\rightarrow J/\psi (\eta_c)$;
the last two diagrams are of  nonfactorizable topology.  With the meson wave functions, Sudakov factors and the six-quark hard subamplitude,
after a straightforward calculation employing the pQCD formalism of Eq.\ref{eq:factorization}, we can get the  explicit expressions of the amplitude
in Fig.\ref{fig:jpv}, which are listed in Appendix \ref{amplitudes}.

The total decay amplitude for  the $B_c\rightarrow (J/\psi, \eta_c)(P, V)$ can be given by
\begin{eqnarray}\label{eq:amplitude1}
\mathcal {A}(B_c\rightarrow (J/\psi, \eta_c)(P, V))&=&V_{cb}^*V_{ud(s)}[(C_2+\frac{1}{3}C_1)\mathcal {F}_{e}+C_1\mathcal {M}_{e}].
\end{eqnarray}
Here, the wilson coefficients $C_{1,2}$ are actually convoluted
with the amplitudes $\mathcal {F}_{e}$ and $\mathcal {M}_{e}$.
Note that the $B_c\rightarrow J/\psi V$ decays  contain more amplitudes associated with   three different polarizations,
 one longitudinal and two transverse for the two vector mesons,  possible.
The amplitude can be decomposed as
\begin{eqnarray}
\mathcal {A}=\mathcal {A}^L+\mathcal {A}^N \epsilon_{2}^T \cdot \epsilon_{3}^T +
i \mathcal {A}^T \epsilon_{\alpha\beta\rho\sigma}n^{\alpha}v^{\beta}\epsilon_{2}^{T \rho}\epsilon_{3}^{T \sigma},
\end{eqnarray}
where  $\epsilon_2^T, \epsilon_3^T$ are the transverse polarization vectors for the two vector  mesons, respectively.
$\mathcal {A}^L$ corresponds to the contributions of  longitudinal polarization; $\mathcal {A}^N$ and $\mathcal {A}^T$
corresponds to the contributions of  normal  and transverse  polarization, respectively,
and the total amplitudes $\mathcal {A}^{L,N,T}$ have the same structures as Eq.(\ref{eq:amplitude1}).
The factorization formulas for the longitudinal, normal and transverse polarizations  are
all listed in Appendix A\ref{amplitudes}.

\subsection{ $B_c\rightarrow (J/\psi, \eta_c)D_{(s)}^{(*)}$ decays}

The effective Hamiltonian for the
flavor-changing $b\rightarrow q'$ transition is given by
\begin{eqnarray}
\mathcal {H}_{eff}=\frac{G_F}{\sqrt{2}}\{(V^*_{cb}V_{cq'}C_1(\mu)O_1(\mu)+C_2(\mu)O_2(\mu))- V^*_{tb}V_{tq'}\Sigma_{i=3}^{10}C_i(\mu)O_i(\mu)\}
\end{eqnarray}
with $q'=d,s$. The functions $Q_i(i=1,2,...,10)$ are the
local four-quark operators:
\begin{enumerate}
  \item  tree operators
\begin{eqnarray}
O_{1}(\mu)&=&\bar{b}_{\alpha}\gamma^{\mu}(1-\gamma_5)c_{\beta} \otimes \bar{c_{\alpha}}\gamma_{\mu}(1-\gamma_5)q'_{\beta},\nonumber\\
O_{2}(\mu)&=&\bar{b}_{\alpha}\gamma^{\mu}(1-\gamma_5)c_{\alpha}\otimes \bar{c}_{\beta}\gamma_{\mu}(1-\gamma_5)q'_{\beta},
\end{eqnarray}
  \item  QCD penguin operators
  \begin{eqnarray}
O_{3}(\mu)&=&\bar{b}_{\alpha}\gamma^{\mu}(1-\gamma_5)q'_{\alpha} \otimes \sum_q \bar{q_{\beta}}\gamma_{\mu}(1-\gamma_5)q_{\beta},\nonumber\\
O_{4}(\mu)&=&\bar{b}_{\alpha}\gamma^{\mu}(1-\gamma_5)q'_{\beta} \otimes \sum_q \bar{q_{\alpha}}\gamma_{\mu}(1-\gamma_5)q_{\beta},\nonumber\\
O_{5}(\mu)&=&\bar{b}_{\alpha}\gamma^{\mu}(1-\gamma_5)q'_{\alpha} \otimes \sum_q \bar{q_{\beta}}\gamma_{\mu}(1+\gamma_5)q_{\beta},\nonumber\\
O_{6}(\mu)&=&\bar{b}_{\alpha}\gamma^{\mu}(1-\gamma_5)q'_{\beta} \otimes \sum_q \bar{q_{\alpha}}\gamma_{\mu}(1+\gamma_5)q_{\beta},
\end{eqnarray}
  \item electroweak penguin operators
  \begin{eqnarray}
O_{7}(\mu)&=&\frac{3}{2}\bar{b}_{\alpha}\gamma^{\mu}(1-\gamma_5)q'_{\alpha} \otimes \sum_q e_q\bar{q_{\beta}}\gamma_{\mu}(1+\gamma_5)q_{\beta},\nonumber\\
O_{8}(\mu)&=&\frac{3}{2}\bar{b}_{\alpha}\gamma^{\mu}(1-\gamma_5)q'_{\beta} \otimes \sum_q e_q\bar{q_{\alpha}}\gamma_{\mu}(1+\gamma_5)q_{\beta},\nonumber\\
O_{9}(\mu)&=&\frac{3}{2}\bar{b}_{\alpha}\gamma^{\mu}(1-\gamma_5)q'_{\alpha} \otimes \sum_q e_q\bar{q_{\beta}}\gamma_{\mu}(1-\gamma_5)q_{\beta},\nonumber\\
O_{10}(\mu)&=&\frac{3}{2}\bar{b}_{\alpha}\gamma^{\mu}(1-\gamma_5)q'_{\beta} \otimes \sum_q e_q\bar{q_{\alpha}}\gamma_{\mu}(1-\gamma_5)q_{\beta}.
\end{eqnarray}
\end{enumerate}
 The sum over $q$ runs
over the quark fields that are active at the scale $\mu=O(m_b)$,
 i.e. $q=(u,d,s,c,b)$.


\begin{figure}[!htbh]
\begin{center}
\vspace{-1cm}\centerline{\epsfxsize=10 cm \epsffile{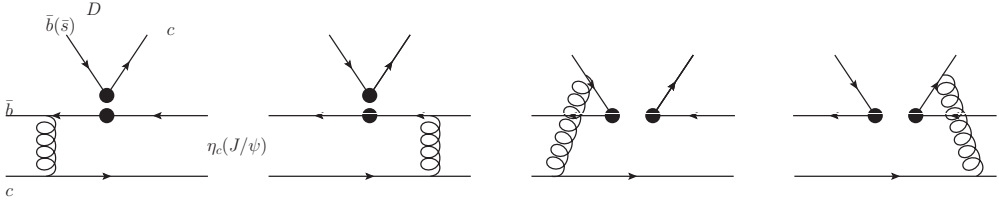}}
\vspace{-5.5cm}\caption{Color-favored  diagrams contributing to the Feynman diagrams for  $B_c\rightarrow (J/\psi, \eta_c)D_{(s)}^{(*)}$ decays.}
\label{fig:jDallow}
 \end{center}
\end{figure}
 \begin{figure}[!htbh]
\begin{center}
\vspace{-1cm} \centerline{\epsfxsize=10 cm \epsffile{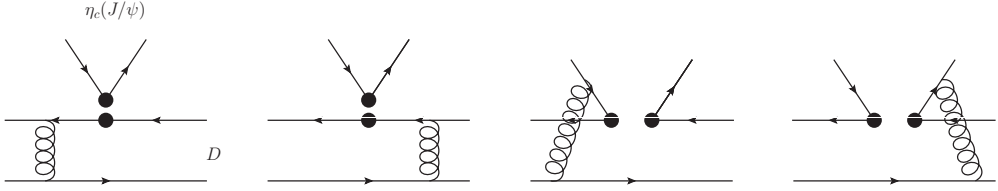}}
\vspace{-5.5cm} \caption{Color-suppressed diagrams contributing to the Feynman diagrams for  $B_c\rightarrow (J/\psi, \eta_c)D_{(s)}^{(*)}$ decays.}
 \label{fig:jDsupp}
 \end{center}
\end{figure}
\begin{figure}[!htbh]
\begin{center}
\vspace{-1cm} \centerline{\epsfxsize=11 cm \epsffile{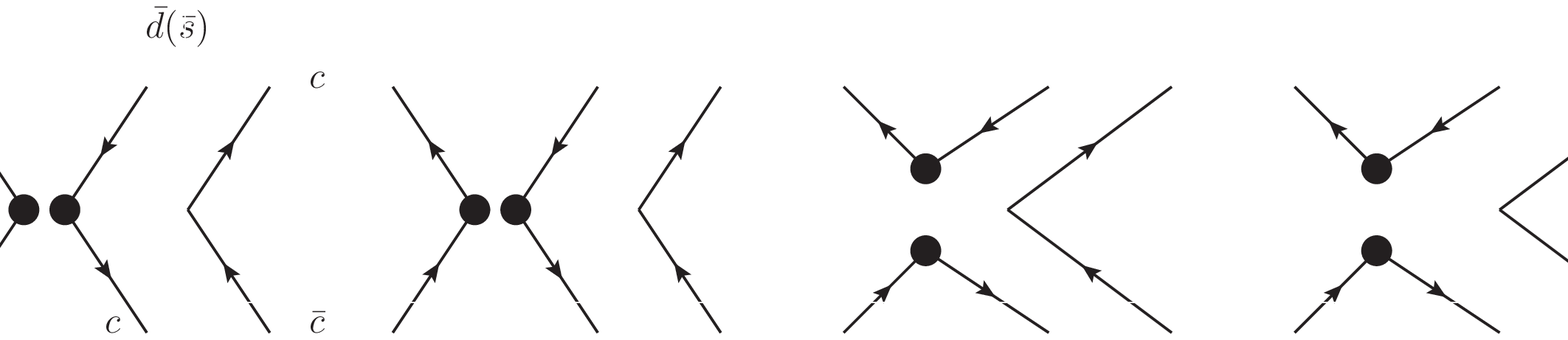}}
\vspace{-5.5cm} \caption{Annihilation diagrams contributing to the Feynman diagrams for  $B_c\rightarrow (J/\psi, \eta_c)D_{(s)}^{(*)}$ decays.}
 \label{fig:jDann}
 \end{center}
\end{figure}

There are 12  Feynman diagrams contributing to $B_c\rightarrow (J/\psi, \eta_c)D_{(s)}^{(*)}$ decays
at leading order. They
involve three types:  color-favored  diagrams (we mark this kind of contribution with the subscript $f$ ) shown in Fig.\ref{fig:jDallow}, color-suppressed diagrams  (marked with $s$)  shown in Fig.\ref{fig:jDsupp}  and annihilation diagrams  (marked with $a$) shown in Fig.\ref{fig:jDann}. Each type is classified into factorizable diagrams,
where hard gluon connects the quarks in the same meson,
and nonfactorizable diagrams, where hard gluon attaches
the quarks in two different mesons. We also
show the calculated formulas of each diagram for different channels in  Appendix  A \ref{amplitudes}.
The total decay amplitude for decay is given as
\begin{eqnarray}\label{eq:amplitude2}
\mathcal {A}(B_c\rightarrow (J/\psi, \eta_c)D_{(s)}^{(*)})&=&V_{cb}^*V_{cd}[(C_2+\frac{1}{3}C_1)\mathcal {F}_{f}^{LL}+C_1\mathcal {M}_{f}^{LL}+\nonumber\\&&(C_1+\frac{1}{3}C_2)\mathcal {F}_{s}^{LL}+C_2\mathcal {M}_{s}^{LL}+(C_2+\frac{1}{3}C_1)\mathcal {F}_{a}^{LL}+C_1\mathcal {M}_{a}^{LL}]
\nonumber\\&&-V_{tb}^*V_{td}[(C_4+\frac{1}{3}C_3+C_{10}+\frac{1}{3}C_9)\mathcal {F}_{f}^{LL}+(C_3+C_9)\mathcal {M}_{f}^{LL}\nonumber\\&&+(C_3+\frac{1}{3}C_4+C_{9}+\frac{1}{3}C_{10})\mathcal {F}_{s}^{LL}+(C_4+C_{10})\mathcal {M}_{s}^{LL}\nonumber\\&&+(C_4+\frac{1}{3}C_3+C_{10}+\frac{1}{3}C_9)\mathcal {F}_{a}^{LL}+(C_3+C_9)\mathcal {M}_{a}^{LL}\nonumber\\&&+(C_6+\frac{1}{3}C_5+C_{8}+\frac{1}{3}C_7)\mathcal {F}_{f}^{SP}+(C_6+C_8)\mathcal {M}_{s}^{SP}\nonumber\\&&+(C_5+\frac{1}{3}C_6+C_{7}+\frac{1}{3}C_8)\mathcal {F}_{s}^{LR}+(C_5+C_7)\mathcal {M}_{f}^{LR}\nonumber\\&&+(C_6+\frac{1}{3}C_5+C_{8}+\frac{1}{3}C_7)\mathcal {F}_{a}^{SP}+(C_5+C_7)\mathcal {M}_{a}^{LR}].
\end{eqnarray}
Note that the amplitude $\mathcal {F}_{f}^{SP}$ from the operators $O_{5-8}$ vanishes when a vector meson ($D^{*}_{(s)}$) is emitted from the weak vertex,
 because neither the scalar nor the pseudoscalar density gives contributions to the vector meson production, i.e. $\langle D^{*}_{(s)}|S+P|0\rangle=0$.


\section{NUMERICAL RESULTS} \label{sec:result}
We now use the method previously illustrated to estimate the physical observables (such as transiiton form factors, branching ratios, transverse polarization fractions and direct CP violations) of the  considered $B_c$ decays.
For numerical calculation,  some input parameters needed in the pQCD calculation are listed in Table \ref{tab:constant}, while the input wave functions and various parameters of the light  vector and pseudoscalar mesons are shown in the corresponding paper \cite{prd074008}.
If not specified explicitly,
we will take their central values as the default input.
\begin{table}
\caption{The decay constants of mesons are from \cite{epjc60107,prd074019}, while other parameters are adopted in PDG \cite{pdg} in  our numerical calculation.  }
\label{tab:constant}
\begin{tabular*}{14.5cm}{@{\extracolsep{\fill}}l|ccccc}
  \hline\hline
\textbf{Mass(\text{GeV})} & $M_{W}=80.399$   & $M_{B_c}=6.277$ & $m_{b}=4.2$& $m_{c}=1.27$ & $m_{J/\psi}=3.097$\\[1ex]
&$m_{\eta_c}=2.981$&$m_D=1.870$&$m_{D^*}=2.010$&$m_{D_s}=1.968$&$m_{D_s^*}=2.112$\\[1ex]
\hline
\end{tabular*}
\begin{tabular*}{14.5cm}{@{\extracolsep{\fill}}l|ccc}
  \hline
\multirow{2}{*}{{\textbf{CKM}}} & $|V_{cb}|=(40.9\pm1.1)\times 10^{-3}$
 & $|V_{ud}|=0.97425\pm0.00022$ & $|V_{us}|=0.2252\pm0.0009$\\[1ex]
&$|V_{ub}|=(4.15\pm0.49)\times 10^{-3}$ & $|V_{cd}|=0.230\pm0.011$
& $|V_{cs}|=1.006\pm0.023$  \\[1ex]
\hline
\end{tabular*}
\begin{tabular*}{14.5cm}{@{\extracolsep{\fill}}l|ccccc}
\hline
\textbf{Decay constants(MeV)} & $f_{B_c}=489\pm4\pm3$ & $f_{J/\psi}=405\pm 14$& $f_{\eta_c}=420\pm 50$\\[1ex]
&$f_{D}=206.7 \pm 8.9$ & $f_{D_s}=257.5 \pm 6.1$\\[1ex]
\hline
\end{tabular*}
\begin{tabular*}{14.5cm}{@{\extracolsep{\fill}}l|l}
\hline
\textbf{Lifetime} & $\tau_{B_c}=0.453\times 10^{-12}\text{s}$\\[1ex]
\hline\hline
\end{tabular*}
\end{table}
\subsection{$B_c\rightarrow \eta_c,J/\psi$ form factors}
The diagrams (a) and (b) in Fig.\ref{fig:jpv} give the contribution for $B_c\rightarrow \eta_c,J/\psi$ transition form factor
 at the maximally recoiling point ($q^2=0$). Our predictions of the   form factors are collected in Table \ref{tab:formfactor} compared with the results from other models.
  The first kind of uncertainties
 is from  the  uncertainty in  the hadronic parameters:  $\omega_B=0.60\pm0.05$ \cite{prd074019} for
$B_c$ meson and $\omega=0.60\pm0.1$ \cite{epjc60107}for
$J/\psi(\eta_c)$ meson, while the second kind
of uncertainties is from those in decay constants of
the $B_c$ meson and the  charmonium  meson, which are given in Table \ref{tab:constant}.
We find both $A_0^{B_c\rightarrow J/\psi}$ and $F_0^{B_c\rightarrow \eta_c}$  decrease with increasing shape parameters $\omega_B$ and $\omega$.
The  former are more sensitive to the shape parameters than the decays constants, while the latter is just the reverse.
Since the uncertainties from decay constant of $\eta_c$ meson are large,
the  relevant uncertainties to $F_0^{B_c\rightarrow \eta_c}$ are also large.
We can see that the  $B_c\rightarrow \eta_c,J/\psi$ transition
form factors  are larger than  those of $B_c\rightarrow D_{(s)}^{(*)}$ in our previous study \cite{prd074019} under
the perturbative QCD approach. As it is well known, compared with the $D$ meson, the  $J/\psi(\eta_c)$ meson is heavier, and its velocity is  lower in the rest frame of the
$B_c$ meson. The overlap between the initial and final state wave functions becomes larger,
which certainly induces larger form factors.
\begin{table}
\caption{The form factors for $F_0^{B_c\rightarrow \eta_c}$ and $A_0^{B_c\rightarrow J/\psi}$ at $q^2=0$ evaluated in the literature.
The  uncertainties are from the  hadronic parameters and the decay constants, respectively. For comparison, we also cite the theoretical estimates of other models.}
\label{tab:formfactor}
\begin{tabular}[t]{l|c|c|c|c|c|c|c|c}
\hline\hline
 & This work & SDY\cite{epjc60107}\footnotemark[1]& Kiselev \cite{npb569473}& IKP \cite{prd074010} & WSL \cite{prd054012}
 & HZ \cite{epjc51833} & DSV \cite{jpg085002} &EFG\cite{prd094020}\\ \hline
$F_0^{B_c\rightarrow \eta_c}$     & $0.72^{+0.10+0.08}_{-0.08-0.09}$ &$0.66\sim 0.79$ &0.66  &0.76   &0.61 &0.87 &0.58&0.47\\
$A_0^{B_c\rightarrow J/\psi}$& $0.64^{+0.08+0.02}_{-0.07-0.02}$&$0.65\sim 0.77$ &0.60   &0.69 &0.53&0.27  &0.58&0.40\\
\hline\hline
\end{tabular}
\footnotetext[1]{We quote  the result with the charmonium wave function for a harmonic-oscillator potential}
\end{table}

The $B_c\rightarrow J/\psi,\eta_c$ transition form factors have been widely studied in
many theoretical frameworks, which are also collected in Table.\ref{tab:formfactor}. Most of our results are found to be comparable to those of \cite{epjc60107,npb569473,prd074010,prd054012,epjc51833,jpg085002}, whereas the form factor $A_0^{B_c\rightarrow J/\psi}$ in Ref. \cite{epjc51833} is typically smaller, which can be discriminated by the future
LHC experiments.

\subsection{Branching ratios}
\begin{table}
\caption {Branching ratios ($10^{-3}$) for $B_c\rightarrow (J/\psi, \eta_c)(P, V)$, together with results from other models.
The errors for these entries correspond to the uncertainties in  hadronic shape parameters, from the decay constants, and
the scale dependence, respectively.}
\label{tab:branching1}
\begin{tabular}[t]{l|c|c|c|c|c|c|c|c|c}
\hline\hline
 Channels & This work & \cite{prd034008} & \cite{prd094020}& \cite{prd3399}& \cite{npb585353}
 &\cite{prd054024} & \cite{prd094028}& \cite{prd034012}& \cite{prd014019}
  \\ \hline
$B^+_c\rightarrow J/\psi K^+$     &$0.19^{+0.04+0.02+0.02}_{-0.04-0.02-0.01}$   &0.22 &0.05&0.16 &0.11&0.13 &0.11&0.07&0.08 \\
$B^+_c\rightarrow J/\psi K^{*+}$    &$0.48^{+0.09+0.04+0.05}_{-0.08-0.03-0.03}$ &0.43 &0.10 &0.35 &0.22&0.28 &0.09& 0.2 &0.18\\
$B^+_c\rightarrow \eta_c K^+$     &$0.24^{+0.04+0.07+0.02}_{-0.05-0.06-0.01}$   &0.38 &0.07&0.17 &0.13&0.15 &0.03& 0.02 &0.11 \\
$B^+_c\rightarrow \eta_c K^{*+}$    &$0.57^{+0.10+0.11+0.06}_{-0.08-0.09-0.03}$ &0.77 &0.11 &0.31 &0.20&0.25 &0.06& 0.04 &0.18\\
\hline
$B^+_c\rightarrow J/\psi\pi^+$       & $2.33^{+0.63+0.16+0.48}_{-0.58-0.16-0.12}$&2.91 &0.61 &2.1  &1.3&1.7 &0.34 & 1.3 &1.1 \\
$B^+_c\rightarrow J/\psi\rho^+$     & $8.20^{+1.49+0.58+0.50}_{-1.28-0.56-0.62}$&8.08 &1.6 &6.5  &4.0&4.9 &1.8  &3.7   &3.1\\
$B^+_c\rightarrow \eta_c\pi^+$       & $2.98^{+0.84+0.75+0.52}_{-0.79-0.67-0.14}$&5.19 &0.85 &2.2  &2.0&1.9 &0.34 & 0.26 &1.4\\
$B^+_c\rightarrow \eta_c\rho^+$     & $9.83^{+1.38+2.48+1.74}_{-1.29-2.20-0.47}$&14.5 &2.1 &5.9  &4.2&4.5 &1.06 &0.67  &3.3\\
\hline\hline
\end{tabular}
\end{table}
\begin{table}
\caption {Branching ratios ($10^{-3}$) for $B_c\rightarrow (J/\psi, \eta_c)D^{(*)}_{(s)}$, together with results from other models.
The errors for these entries correspond to the uncertainties in  hadronic shape parameters, from the decay constants, and
the scale dependence, respectively.}
\label{tab:branching2}
\begin{tabular}[t]{l|c|c|c|c|c|c|c|c}
\hline\hline
 Channels & This work &\cite{prd3399}& \cite{npb585353} & \cite{prd054024} & \cite{prd094028} &  \cite{prd034012}
& \cite{prd014019} &\cite{jhep06015}  \\
 \hline
$B^+_c\rightarrow J/\psi D^+$      & $0.28^{+0.04+0.02+0.06}_{-0.04-0.02-0.02}$  &0.009&0.09&0.15  &0.04 &0.13  &0.09 &0.09            \\
$B^+_c\rightarrow J/\psi D^{*+}$   & $0.67^{+0.11+0.05+0.15}_{-0.09-0.05-0.05}$  &-&0.28&0.45  &0.18 &0.19  &0.28  &0.28          \\
$B^+_c\rightarrow \eta_c D^+$      & $0.44^{+0.07+0.11+0.07}_{-0.06-0.09-0.03}$  &0.012&0.15&0.19  &0.06 &0.05  &0.14  &0.10             \\
$B^+_c\rightarrow \eta_c D^{*+}$   & $0.58^{+0.10+0.15+0.11}_{-0.08-0.13-0.04}$  &0.010&0.10&0.19  &0.07 &0.02   &0.13 &0.10            \\
\hline
$B^+_c\rightarrow J/\psi D_s^+$    &$8.05^{+1.39+0.57+1.66}_{-1.18-0.54-0.36}$   &0.41&1.7&3.4   &1.15  &3.4  &1.5&2.2          \\
$B^+_c\rightarrow J/\psi D^{*+}_s$ &$20.45^{+4.35+1.44+4.50}_{-4.05-1.39-3.00}$  &-&6.7&9.7   &4.4   &5.9  &5.5&6.0        \\
$B^+_c\rightarrow \eta_c D_s^+$    &$12.32^{+2.06+3.11+2.03}_{-1.79-2.76-1.01}$  &0.54&2.8&4.4   &1.79  &5    &2.6&2.5          \\
$B^+_c\rightarrow \eta_c D_s^{*+}$ &$16.54^{+2.72+4.17+3.09}_{-2.34-3.70-1.70}$  &0.44&2.7&3.7   &1.49  &0.38 &2.4&2.0         \\
\hline\hline
\end{tabular}
\end{table}
The branching ratios in the $B_c$ meson rest frame can be written as
\begin{eqnarray}
\mathcal {BR}(B_c\rightarrow(J/\psi, \eta_c)D_{(s)}^{(*)})&=&\frac{G_F^2\tau_{B_c}}{32\pi M_B}\sqrt{1-(r_D-r_{J/\psi(\eta_c)})^2}\sqrt{1-(r_D+r_{J/\psi(\eta_c)})^2}|\mathcal {A}|^2,\nonumber\\
\mathcal {BR}(B_c\rightarrow(J/\psi, \eta_c)(P, V))&=&\frac{G_F^2\tau_{B_c}}{32\pi M_B}(1-r_{J/\psi(\eta_c)}^2)|\mathcal {A}|^2,
\end{eqnarray}
where the mass ratios $r_i $ and the decay amplitudes $\mathcal {A}$ for each channel have been given explicitly in Appendix \ref{amplitudes}.
Our numerical results of branching ratios for $B_c\rightarrow(J/\psi, \eta_c)(P, V)$ and $B_c\rightarrow(J/\psi, \eta_c)D_{(s)}^{(*)}$ decays are listed in Tables \ref{tab:branching1} and Table \ref{tab:branching2}, respectively.  The  first two  errors  are the same as 
for form factors in Table \ref{tab:formfactor}, while the third error arises from the
hard scale $t$ varying from $0.75t$ to $1.25t$, which characterizes
the size of next-to-leading order (NLO) QCD contributions. We can see  the branching ratios are sensitive to the choice of the hadronic parameters $\omega_B$
and $\omega$, the combined uncertainties from them are about 20\%. In addtion,  the uncertainties from the decay constants except for $f_{\eta_c}$ are small.
However, for $B_c\rightarrow (J/\psi,\eta_c)D^{(*)}_{(s)}$ decays, the uncertainties from the hard scale $t$ is large as shown in Table \ref{tab:branching2},
which means the next-to-leading order contributions may be important for this decay mode.
It reflects that the energy release in this type decay may be low for pQCD to play.
The  similar situation also exists in $B_c\rightarrow BP,BV$
\cite{prd114019} and
$B\rightarrow (J/\psi,\eta_c)D^{(*)}$ \cite{liying} decays. In a recent paper \cite{prd034008}, the authors have performed the $B_c$ meson exclusive decays to
S-wave charmonia and light pseudoscalar or vector mesons at the next-to-leading order (NLO) in the QCD coupling. The  NLO corrections to
$B_c$ decays under the pQCD framework are still missing, thus
 beyond the scope of this paper.
 From  Tables \ref{tab:branching1} and  \ref{tab:branching2}, we can see
 the former four processes  have a relatively small branching ratio ($10^{-4}$)
owing to the CKM factor suppression, while the branching ratios of other processes are comparatively large ($10^{-3}\sim 10^{-2}$) due to the CKM
factor enhancement. The large branching ratio and the clear signals of the final states make their measurement easy at the LHCb experiments.

For comparison, we also cite other theoretical results \cite{prd034008,prd094020,prd3399,npb585353,prd054024,prd094028,prd034012,prd014019,jhep06015} for the considered decays in Tables \ref{tab:branching1} and  \ref{tab:branching2}. In general, the results of the various model calculations are of the same order of magnitude for most channels,
while our predictions are  larger
than those of other approaches. The difference may due to at least two reasons:  First, the calculations  in Refs.\cite{prd094020,prd3399,npb585353,prd054024,prd094028,prd034012,prd014019,jhep06015}  use the same naive factorization
approximation in which the form factors are important input parameters,
 smaller form factors  always result in  the smaller branching ratios.
Second, in pQCD framework, the  nonfactorizable
contribution is considered, which is absent in traditional naive factorization. A constructive interference between
the  nonfactorizable  contribution and the  factorizable contribution will enhance our results. From Table. \ref{tab:branching1}, we can see that
our predicted branching ratios are comparable with \cite{prd034008} which also include the nonfactorizable  contribution.
Since the charmonium decays  dominate to the $b\rightarrow c,u$ induced $B_c$ decays,  summing up all the
branching ratios in   Tables \ref{tab:branching1} and  \ref{tab:branching2} one obtains a total branching ratio of
$10\%$ which has to be compared with the $20\%$ expected
for  the $b \rightarrow c,u$ contribution to the total rate \cite{iiba}, this leaves
plenty of room for the $B_c$ meson to charmonium semileptonic, excited charmonuim meson  and nonresonant multibody decays.

The two decays $B_c\rightarrow J/\psi \pi, J/\psi K$  have identical topology and similar kinematic
properties, as shown in Fig. \ref{fig:jpv}. In the limit of $SU(3)$ flavor symmetry, the ratio of branching fractions
$\mathcal {BR}(B_c\rightarrow J/\psi K)/\mathcal {BR}(B_c\rightarrow J/\psi \pi)$
is dominated by the ratio of the relevant  CKM matrix
elements $|V_{us}/V_{ud}|^2$. After including the decay constants  $f_{K(\pi)}$,
the ratio is enhanced. With the input parameters in Table \ref{tab:constant}, the expected ratio is 0.080, which is very close to
our prediction 0.082.  It means that the dominant contributions to the branching ratios come from the factorizable topology, while
the nonfactorizable contribution is suppressed by the Wilson coefficient $C_1$ (see Eq.\ref{eq:amplitude1}).
Recently, the LHCb collaboration has measured this  ratio to be $0.069 \pm 0.019 \pm 0.005$
which is compatible  with our pQCD prediction. For $B_c\rightarrow \eta_c \pi, \eta_c K$ decays, our result of $\mathcal {BR}(B_c\rightarrow \eta_c K)/\mathcal {BR}(B_c\rightarrow \eta_c \pi)$ is 0.081, which will be tested by the forthcoming experiments.

Due to $m_{J/\psi} > m_{\eta_c}$ and the orbital angular
momentum of the final states $J/\psi M$ are larger than that of $\eta_c M$
, the phase space for $B_c\rightarrow J/\psi M$ decay is tighter than
that for $B_c\rightarrow \eta_c M$ decay. Therefore, with the same input,
the branching ratios for $B_c\rightarrow J/\psi M$ and $B_c\rightarrow \eta_c M$ decays
have the following hierarchy
\begin{eqnarray}\label{eq:relation}
\mathcal {BR}(B_c\rightarrow J/\psi M)<\mathcal {BR}(B_c\rightarrow \eta_c M).
\end{eqnarray}
However, for $B_c\rightarrow J/\psi D^{*}_{(s)}$ decays, the transverse polarization amplitude contributes to the branching ratio as large as
the longitudinal polarization amplitude, which spoils the  hierarchy relation in Eq.\ref{eq:relation}.

It may be noted that the $B_c\rightarrow (J/\psi,\eta_c)D^{(*)}_{(s)}$ decays involve contributions
from  the color-favored, color-suppressed and  weak annihilation diagrams. It is expected that
the color-favored factorizable amplitude $\mathcal {F}_{f}^{LL}$ dominates in Eq.\ref{eq:amplitude2}.
The color-suppressed nonfactorizable amplitude  $\mathcal {M}_{s}^{LL}$
and the annihilation amplitude $\mathcal {F}_{a}^{LL}$, are enhanced by the large Wilson coefficient $C_2$ and $C_2+\frac{1}{3}C_1$, respectively.
However,  the contribution from $\mathcal {F}_{a}^{LL}$ are highly power suppressed due to a big cancellation between the first two factorizable annihilation diagrams in  Fig. \ref{fig:jDann}. Our numerical analysis shows that $(C_2+\frac{1}{3}C_1)\mathcal {F}_{a}^{LL}/(C_2+\frac{1}{3}C_1)\mathcal {F}_{f}^{LL}\sim 1\%$ and $C_2\mathcal {M}_{s}^{LL}/(C_2+\frac{1}{3}C_1)\mathcal {F}_{f}^{LL}\sim 10\%$.
The interferences between $\mathcal {F}_{f}^{LL}$ and $\mathcal {M}_{s}^{LL}$ are constructive,
  while, the existing experimental data favor constructive interference in the $B$ meson decays \cite{ppnp81}.
The predicted branching ratios of these modes would provide an interesting test of interference between the color-favored and color-suppressed $B_c$
decays.
Experimentally, the available measurements of the considered $B_c$ decay  are as follows \cite{prd112012}
\begin{eqnarray}
\frac{\mathcal {BR}(B_c\rightarrow J/\psi D_s)}{\mathcal {BR}(B_c\rightarrow J/\psi \pi)}&=&2.90\pm0.57\pm0.24,\nonumber\\
\frac{\mathcal {BR}(B_c\rightarrow J/\psi D^*_s)}{\mathcal {BR}(B_c\rightarrow J/\psi D_s)}&=&2.37\pm0.56\pm0.10,
\end{eqnarray}
which is consistent with our predictions,
\begin{eqnarray}
\frac{\mathcal {BR}(B_c\rightarrow J/\psi D_s)}{\mathcal {BR}(B_c\rightarrow J/\psi \pi)}&=&3.45^{+0.49}_{-0.17},\nonumber\\
\frac{\mathcal {BR}(B_c\rightarrow J/\psi D^*_s)}{\mathcal {BR}(B_c\rightarrow J/\psi D_s)}&=&2.54_{-0.21}^{+0.07}.
\end{eqnarray}
\subsection{Transverse polarization fractions}
For the $B_c$ decays to two vector mesons, the decay amplitudes $\mathcal {A}$ are defined in the helicity basis
\begin{eqnarray}
\mathcal {A}=\sum_{i=0,+,-}|\mathcal {A}_i|^2,
\end{eqnarray}
where the helicity amplitudes $\mathcal {A}_i$ have the following relationships with $\mathcal {A}^{L,N,T}$£º
\begin{eqnarray}
\mathcal {A}_0=\mathcal {A}^{L},\quad \mathcal {A}_{\pm}=\mathcal {A}^{N}\pm \mathcal {A}^{T}.
\end{eqnarray}
We also calculate the transverse polarization fractions $\mathcal {R}_{T}$ of the $B_c\rightarrow J/\psi (\rho,K^*,D_{(s)}^*)$
decays, with the definition given by
\begin{eqnarray}
\mathcal {R}_T=\frac{|\mathcal {A}_{+}|^2+|\mathcal {A}_{-}|^2}{|\mathcal {A}_{0}|^2+|\mathcal {A}_{+}|^2+|\mathcal {A}_{-}|^2}.
\end{eqnarray}
According to the power counting rules in the factorization assumption,
 the longitudinal polarization dominates the decay ratios and the transverse polarizations are
suppressed \cite{1979} due to the helicity flips of the quark in the final state hadrons.
Our predictions for the transverse polarization fractions of the tree-dominated $B_c\rightarrow J/\psi V$ decays are given in Table \ref{tab:ratio}.
These results have the following pattern
\begin{eqnarray}
\mathcal {R}_T(J/\psi \rho)<\mathcal {R}_T(J/\psi K^*)<\mathcal {R}_T( J/\psi D^*)<\mathcal {R}_T( J/\psi D^*_s).
\end{eqnarray}
It can be simply
understood by means of kinematics in the heavy-quark  limit.
The transverse polarization fractions $\mathcal {R}_{T}$ of the $B_c\rightarrow J/\psi (\rho,K^*,D^*,D_s^*)$
modes increase as the masses of the mesons $\rho,K^*,D^*,D_s^*$
emitted from the weak vertex increase. This is similar to the case of $B^0\rightarrow (\rho^+,D^{*+},D_s^{*+})D^{*-}$ and
$B^+\rightarrow (\rho^+,D^{*+},D_s^{*+})\rho^0$ \cite{prd054025}. From Table  \ref{tab:ratio},
the  modes $B_c\rightarrow J/\psi (\rho, K^*)$
are indeed longitudinal polarization dominant, since the two transverse amplitudes are
down by a power of $r_{J/\psi}$ or $r_{v}$ comparing with the longitudinal amplitudes.
 However, for  $B_c\rightarrow J/\psi D^*, J/\psi D_s^*$ decays, the transverse polarization fractions can reach  46\% and 48\%, respectively.
Several reasons are given in order. First, the mass ratio $r_D$ for $D^*$ meson is  about 2-3 times larger than the  $r_v$ for light vector meson,
 which enhances the color-favored transverse amplitude $\mathcal {F}^{LL,T}_f$ and the normal amplitude $\mathcal {F}^{LL,N}_f$.
 Second, the annihilation contribution of operator $O_6$  ($\mathcal {F}^{SP,N(T)}_a$)  is chirally enhanced in pQCD
approach \cite{0606094}. Third, the transverse polarization of the nonfactorizable color-suppressed diagrams in Figs. \ref{fig:jDsupp}(c) and \ref{fig:jDsupp}(d) does not encounter helicity flip suppression \cite{prd074019}. The combined effect above enhances the transverse polarization fractions of the
$B_c\rightarrow J/\psi D^*_{(s)}$ decays.
Therefore, the above predictions on the transverse polarization fractions are reasonable in pQCD
framework and comparable with the relativistic independent quark model (RIQM) \cite{prd094028}.
The measurement of polarization fraction
for $B_c\rightarrow J/\psi D^*_s$ decay by the LHCb measurement \cite{prd112012} is
\begin{eqnarray}
\mathcal {R}_T(J/\psi D^*_s)=(52\pm20)\%,
\end{eqnarray}
which is  in good agreement with our result, while
 other predictions can be
tested by the future data.
\begin{table}
\caption{The transverse polarizations fractions ($\%$) for $B_c\rightarrow VV$, together with results from RCQM \cite{prd094028}.
 The errors correspond to the combined uncertainty in the hadronic parameters, decay constants and  the hard scale. }
\label{tab:ratio}
\begin{tabular}[t]{l|c|c|c|c}
\hline\hline
Channels & $B_c\rightarrow J/\psi \rho$ & $B_c\rightarrow J/\psi K^*$
 & $B_c\rightarrow J/\psi D^*$ & $B_c\rightarrow J/\psi D_s^*$\\
\hline
This work &$8^{+2}_{-1}$&$10^{+1}_{-1}$&$46^{+4}_{-3}$&$48^{+4}_{-4}$\\
\hline
RIQM \cite{prd094028}&7 &10&41&43\\
\hline\hline
\end{tabular}
\end{table}



\subsection{The direct CP asymmetries}
Since there is only one kind of CKM phase involved in $B_c$ decaying into   charmonium and a light meson  process, there should be no CP
violation  within the standard model. When the final states are  charmonium and charmed meson, the CP asymmetries arise from the interference between
the penguin diagrams and tree diagrams. The direct CP asymmetry $A^{dir}_{CP}$ for a given mode can be written as
\begin{eqnarray}\label{eq:cpviolation}
A^{dir}_{CP}=\frac{|\mathcal {A}|^2-|\mathcal {\bar{A}}|^2}{|\mathcal {A}|^2+|\mathcal {\bar{A}}|^2},
\end{eqnarray}
where $\mathcal {\bar{A}}$ is the charge conjugate decay amplitude of $\mathcal {A}$, which can be obtained by conjugating the CKM elements in $\mathcal {A}$.
The direct CP asymmetry is tabulated in Table \ref{tab:cpviolation} compared with the results from the
Salpeter method \cite{jhep06015}.
Unlike the  branching ratios, the direct CP asymmetry is not sensitive to the
wave function parameters and CKM factors, since these parameter dependences canceled out in Eq. (\ref{eq:cpviolation}). In addition, the
CKM angles ($\gamma$) uncertainty is quite small ($\sim1\%$). Therefore,
 the theoretical error here is only referred to as the hard scale $t$.
\begin{table}
\caption{The direct CP asymmetry parameters ($10^{-3}$) for $B_c\rightarrow (J/\psi, \eta_c)D_{(s)}^{(*)}$, together with results from the
Salpeter method \cite{jhep06015}.
 The errors arises from the
hard scale t. }
\label{tab:cpviolation}
\begin{tabular}[t]{l|c|c|c|c|c|c|c|c}
\hline\hline
Final stats & $ J/\psi D$ & $ J/\psi D^*$  & $ J/\psi D_s$ & $ J/\psi D_s^*$& $ \eta_c D$ & $ \eta_c D^*$  & $ \eta_c D_s$ & $ \eta_c D_s^*$\\
\hline
This work &$1.5^{+0.6}_{-0.7}$&$12.7^{+4.0}_{-3.1}$&$0.1^{+0.1}_{-0.1}$&$0.7^{+0.2}_{-0.1}$
&$-4.3^{+1.5}_{-2.0}$&$-2.4^{+0.2}_{-0.2}$&$-0.2^{+0.1}_{-0.1}$&$-0.1^{+0.03}_{-0.05}$
\\
\hline
 \cite{jhep06015}&$2.56$&$16.9$&$-0.151$&$-0.972$
&$46.6$&$16.8$&$-2.69$&$-0.965$\\
\hline\hline
\end{tabular}
\end{table}
It can be seen our predictions on direct CP asymmetry parameters of $B_c\rightarrow \eta_c D^{(*)}_{(s)}$ are negative,
while the direct CP asymmetry parameter of the other modes is positive.
The direct CP asymmetry parameters of the processes with a $D$ meson in the final state are generally larger than those with a $D_s$
meson, since in the former processes the penguin diagram contributions are enhanced by the ratio   $\frac{V^*_{tb}V_{td}}{V^*_{cb}V_{cd}}=7.9$,
while in the latter processes,  $\frac{V^*_{tb}V_{ts}}{V^*_{cb}V_{cs}}=0.9$.
However, the penguin amplitudes are still suppressed by the small  Wilson coefficients from penguin operators in both of the two types of mode,
our predictions on direct CP asymmetries are typically smaller in magnitude than \cite{jhep06015} .
From Tables \ref{tab:branching2} and  \ref{tab:cpviolation}, it is easy to see that the decay $B_c\rightharpoonup J/\psi D^{*}$ is helpful to test the CP violating effects due to its large branching ratio and CP asymmetry.

\section{ conclusion}
In the pQCD framework, we have performed a systematic analysis of the two-body nonleptonic decays of  the $B_c$ meson with the final states involving
one $J/\psi(\eta_c)$ meson. Besides the color-favored emission diagrams, the nonfactorizable diagrams and the annihilation diagrams can also be
evaluated in this approach. It is found that the predicted branching ratios range from $10^{-4}$ up to $10^{-2}$, which are easily measured by the running
LHCb in the near future. Our predictions for the  ratios of branching fractions $\frac{\mathcal {BR}(B_c^+\rightarrow J/\Psi D_s^+)}
{\mathcal {BR}(B_c^+\rightarrow J/\Psi \pi^+)}$ ,$\frac{\mathcal {BR}(B_c^+\rightarrow J/\Psi D_s^{*+})}
{\mathcal {BR}(B_c^+\rightarrow J/\Psi D_s^+)}$ and $\frac{\mathcal {BR}(B_c^+\rightarrow J/\Psi K^+)}
{\mathcal {BR}(B_c^+\rightarrow J/\Psi \pi^+)}$ can explain the data perfectly. We also have compared our results with the results of other
studies. In general the results of the various model calculations are of the same order of magnitude while they can
differ by factors of 10 for specific decay modes.
In $B_c$ decaying into one charmonium and one charmed meson process, the CP
violation arises from the interference between the tree diagrams and the penguin diagrams.
We found the direct CP asymmetries of $B_c\rightharpoonup  J/\psi D^{*}$ decays are somewhat large since
the penguin diagrams contributions are enhanced by the CKM factor, which are helpful to test the CP violating effects.
We also find that the transverse polarization contributions in $B_c\rightarrow J/\psi D^*, J/\psi D^*_s$ decays, which
mainly come from the factorizable color-favored  diagrams, the  nonfactorizable color-suppressed diagrams and the chirally enhanced
 annihilation diagrams, are large.

 We also discussed theoretical uncertainties arising from the hadronic parameters, decay constants and hard scale.
The errors in Table \ref{tab:branching1} are dominant by the uncertainties from the hadronic parameters, while in Table \ref{tab:branching2},
the uncertainties from the hard scale are as large as the hadronic parameters due to the included penguin diagram and annihilation diagram. Furthermore, the direct CP asymmetries in Table \ref{tab:cpviolation}
are very sensitive to the scale.
These may suggest that  further studies at the NLO level are required
to improve the accuracy of the theoretical predictions on the charmonium decays of $B_c$ meson.

\begin{acknowledgments}
The authors are grateful to Cai-Dian  L\"{u} and  Junfeng Sun  for helpful discussions.  This work is partially supported by
 National Natural
Science Foundation of China under  Grant No.
11347168 and No. 11347107, and Natural Science Foundation of Hebei Province of China,
  Grant No. A2014209308.

\end{acknowledgments}

\begin{appendix}
\section{the decay amplitudes}\label{amplitudes}


\subsection{Factorization formulas for $B_c\rightarrow \eta_c\rho,\eta_c K^*$}\label{sec:apv}
The decay amplitude of factorizable diagrams in Figs.\ref{fig:jpv}(a) and (b) is
\begin{eqnarray}\label{eq:f}
\mathcal {F}_e&=&2\sqrt{\frac{2}{3}}C_Ff_Bf_{v}\pi M_B^4\sqrt{1-r^2_{\eta _c}}
\int_0^1dx_2\int_0^{\infty}b_1b_2db_1db_2\exp(-\frac{\omega_B^2 b_1^2}{2})\nonumber\\&&\{
[\psi ^s(x_2,b_2)  \left(r_b-2 x_2\right) r_{\eta _c}+\psi ^{\nu }(x_2,b_2)  \left(x_2-2 r_b\right)]E_{ab}(t_a)h(\alpha_e,\beta_a,b_1,b_2)S_t(x_2)
\nonumber\\&&-[\psi ^{\nu }(x_2,b_2)  \left(r_c+ r_{\eta _c}^2\right)-2 \psi^s(x_2,b_2)  r_{\eta _c} ]E_{ab}(t_b)h(\alpha_e,\beta_b,b_2,b_1)S_t(x_1)]\},
 \end{eqnarray}
 with $r_i=m_i/M_B(i=b,c,\eta_c,J/\psi,D,v)$ where $m_i$ are the masses of quark or meson;
$C_F=4/3$ is a color factor; $f_v$ is the decay constant of the
vector   meson, emitted from the weak vertex. The
factorization scales $t_{a,b}$ are chosen as the maximal virtuality
of internal particles in the hard amplitude.
The function $h$ and  $E_{ab}(t)$ are
displayed in Appendix \ref{sec:b}.
 The factor $S_t(x)$ is the jet function from the threshold resummation,
 whose definitions can be found in \cite{epjc45711}. The terms proportional to $r_D^2$ and $r_cr_D$ have been neglected for small values.

 The formula for nonfactorizable in diagrams Fig.\ref{fig:jpv}(c) and (d) is
\begin{eqnarray}\label{eq:me}
\mathcal {M}_e&=&-\frac{8}{3}C_Ff_B\pi M_B^4\sqrt{1-r^2_{\eta _c}}
\int_0^1dx_2dx_3\int_0^{\infty}b_1b_3db_1db_3\phi_{p}^A(x_3)\exp(-\omega_B^2\frac{b_1^2}{2})\nonumber\\&&
\{[\psi ^{\nu }(x_2,b_1)  \left(\left(x_1+2x_2+x_3-2\right) r_{\eta _c}^2+x_1-x_3\right)-\left(x_1+x_2-1\right) \psi ^s(x_2,b_1)  r_{\eta _c}]\nonumber\\&&E_{cd}(t_c)h(\beta_c,\alpha_e,b_3,b_1)
+[\left(x_1+x_2-1\right) \psi ^s(x_2,b_1)  r_{\eta _c}\nonumber\\&&+\psi ^{\nu }(x_2,b_1)  \left((x_3-x_2) r_{\eta _c}^2-2x_1-x_2-x_3+2\right)]E_{cd}(t_d)h(\beta_d,\alpha_e,b_3,b_1)
\}.\nonumber\\
\end{eqnarray}
where
\begin{eqnarray}\label{eq:betaaa}
\alpha_e&=&[x_1+r_{\eta_c}^2(x_2-1)][x_1+x_2-1+r_v^2(1-x_2)]M_B^2, \nonumber\\
\beta_a&=&[r_b^2+(1+r_{\eta_c}^2(x_2-1))(x_2-r_v^2(x_2-1))]M_B^2, \nonumber\\ \beta_b&=&[r_c^2+(r_{\eta_c}^2-x_1)(x_1-1+r_v^2)]M_B^2,\nonumber\\
\beta_c&=&[x_1+x_2-1+r_v^2(1-x_2-x_3)][x_3-x_1-r_{\eta_c}^2(x_2+x_3-1)]M_B^2,\nonumber\\\quad \beta_d&=&[x_1+x_2-1-r_v^2(x_2-x_3)][1-x_1-x_3-r_{\eta_c}^2(x_2-x_3)]M_B^2.
\end{eqnarray}

The corresponding formula  for $B_c\rightarrow \eta_c\pi,\eta_c K$
is similar to Eqs.(\ref{eq:f}) and (\ref{eq:me}), but with the replacement $f_v\rightarrow f_p$, $\phi_{v}\rightarrow \phi_P^A$.
\subsection{Factorization formulas for $B_c\rightarrow J/\psi\rho,J/\psi K^*$}\label{sec:avv}
 We mark $L$, $N$ and $T$ to denote the contributions from longitudinal polarization, normal polarization and transverse polarization, respectively:
\begin{eqnarray}
\mathcal {F}_e^L=&=&2\sqrt{\frac{2}{3}}C_Ff_Bf_{v}\pi M_B^4\sqrt{1-r_{J/\psi}^2}
\int_0^1dx_2\int_0^{\infty}b_1b_2db_1db_2\exp(-\frac{\omega_B^2 b_1^2}{2})\nonumber\\&&\{
[r_{J/\psi} \psi ^t(x_2,b_2) \left(r_b-2 x_2\right)+\psi ^{L }(x_2,b_2) \left(x_2-2r_b\right)]E_{ab}(t_a)h(\alpha_e,\beta_a,b_1,b_2)S_t(x_2)]
\nonumber\\&&-\psi ^{L}(x_2,b_2)[r_{J/\psi}^2 +r_c]E_{ab}(t_b)h(\alpha_e,\beta_b,b_2,b_1)S_t(x_1)]\},
 \end{eqnarray}
\begin{eqnarray}
\mathcal {M}_e^L&=&-\frac{8}{3}C_Ff_B\pi M_B^4\sqrt{1-r_{J/\psi}^2}
\int_0^1dx_2dx_3\int_0^{\infty}b_1b_3db_1db_3\phi_{p}^A(x_3)\exp(-\omega_B^2\frac{b_1^2}{2})\times\nonumber\\&&
\{[\psi ^{L}(x_2,b_1) (r_{J/\psi}^2 -1)\left(x_1-x_3\right)-r_{J/\psi} \left(x_1+x_2-1\right) \psi ^t(x_2,b_1)]\nonumber\\&&E_{cd}(t_c)h(\beta_c,\alpha_e,b_3,b_1)
+[r_{J/\psi} \left(x_1+x_2-1\right) \psi ^t(x_2,b_1)-\nonumber\\&&\Psi ^{L} (x_2,b_1) \left(r_{J/\psi}^2 \left(x_2-x_3\right)+2x_1+x_2+x_3-2\right)]E_{cd}(t_d)h(\beta_d,\alpha_e,b_3,b_1)
\},
\end{eqnarray}
\begin{eqnarray}
\mathcal {F}^N_e&=&2\sqrt{\frac{2}{3}} C_Ff_Bf_{v}\pi M_B^4r_v
\int_0^1dx_2\int_0^{\infty}b_1b_2db_1db_2\exp(-\frac{\omega_B^2 b_1^2}{2})\nonumber\\&&\{
[r_{J/\psi} \psi ^{T}(x_2,b_2) \left(-4 r_b+x_2+1\right)+\left(r_b-2\right) \psi ^V(x_2,b_2)]E_{ab}(t_a)h(\alpha_e,\beta_a,b_1,b_2)S_t(x_2)]
\nonumber\\&&+\psi ^{T}(x_2,b_2)[r_{J/\psi} \left(x_1-1\right) ]
E_{ab}(t_b)h(\alpha_e,\beta_b,b_2,b_1)S_t(x_1)]\},\nonumber\\
 \end{eqnarray}
\begin{eqnarray}
\mathcal {F}^T_e&=&-2\sqrt{\frac{2}{3}} C_Ff_Bf_{v}\pi M_B^4r_v
\int_0^1dx_2\int_0^{\infty}b_1b_2db_1db_2\exp(-\frac{\omega_B^2 b_1^2}{2})\nonumber\\&&\{
[\left(r_b-2\right) \psi ^V(x_2,b_2)-r_{J/\psi} \left(x_2-1\right) \psi ^{T}(x_2,b_2)]E_{ab}(t_a)h(\alpha_e,\beta_a,b_1,b_2)S_t(x_2)]
\nonumber\\&&-\psi ^{T}(x_2,b_2)[r_{J/\psi} \left(x_1-1\right) ]
E_{ab}(t_b)h(\alpha_e,\beta_b,b_2,b_1)S_t(x_1)]\},\nonumber\\
 \end{eqnarray}
\begin{eqnarray}
\mathcal {M}^N_e&=&-\frac{8}{3} C_Ff_B\pi M_B^4 r_v
\int_0^1dx_2dx_3\int_0^{\infty}b_1b_3db_1db_3\phi_{v}(x_3)\exp(-\omega_B^2\frac{b_1^2}{2})\times\nonumber\\&&
\{[2 r_{J/\psi} \left(x_2+x_3-1\right) \psi ^{T}(x_2,b_1) \phi _V^a(x_3)+(x_3-x_1) \psi ^V(x_2,b_1) \phi _V^{\nu }(x_3)]\nonumber\\&&E_{cd}(t_c)h(\beta_c,\alpha_e,b_3,b_1)
-[2 r_{J/\psi} \psi ^{T}(x_2,b_1) \left(\left(x_2-x_3\right) \phi _V^a(x_3)\right.\nonumber\\&&\left.-\left(x_2+x_3-2\right) \phi _V^{\nu }(x_3)\right)+\left(x_1+x_3-1\right) \psi ^V(x_2,b_1) \left(4 \phi _V^a(x_3)+\phi _V^{\nu
   }(x_3)\right)]\nonumber\\&&E_{cd}(t_d)h(\beta_d,\alpha_e,b_3,b_1)
\},
\end{eqnarray}
\begin{eqnarray}
\mathcal {M}^T_e&=&\frac{8}{3} C_Ff_B\pi M_B^4 r_v
\int_0^1dx_2dx_3\int_0^{\infty}b_1b_3db_1db_3\phi_{v}(x_3)\exp(-\omega_B^2\frac{b_1^2}{2})\times\nonumber\\&&
\{[2 r_{J/\psi} \left(x_2-x_3-1\right) \psi ^{T}(x_2,b_1) \phi _V^a(x_3)+(x_3-x_1) \psi ^V(x_2,b_1) \phi _V^{\nu }(x_3)]\nonumber\\&&E_{cd}(t_c)h(\beta_c,\alpha_e,b_3,b_1)
+[2 r_{J/\psi} \psi ^{T}(x_2,b_1) \left(x_2-x_3\right) \phi _V^a(x_3)+\nonumber\\&&\left(x_1+x_3-1\right) \psi ^V(x_2,b_1) \left(4 \phi _V^a(x_3)+\phi _V^{\nu }(x_3)\right)]E_{cd}(t_d)h(\beta_d,\alpha_e,b_3,b_1)
\}.
\end{eqnarray}
where the expression of  $\beta_{a,b,c,d}$ and $\alpha_e$ is the similar to that of Eq. (\ref{eq:betaaa}),
 but with the replacement $r_{\eta_c}\rightarrow r_{J/\psi}$.
For $B_c\rightarrow J/\psi\pi,J/\psi K$ decays, only the longitudinal polarization of $J/\psi$
 will contribute. We can obtain their amplitudes from
the longitudinal polarization amplitudes for the $B_c\rightarrow J/\psi\rho,J/\psi K^*$
decays with the  replacement $f_v\rightarrow f_p$, $\phi_{v}\rightarrow \phi_P^A$.

\subsection{Factorization formulas for $B_c\rightarrow \eta_c D_{(s)}$}\label{sec:apD}
We mark $LL$, $LR$, and $SP$ to denote the contributions
from $(V-A)\otimes(V-A)$, $(V-A)\otimes(V+A)$ and $(S-P)\otimes(S+P)$ operators, respectively:

\begin{eqnarray}
\mathcal {F}^{LL}_f&=&2 \sqrt{\frac{2}{3(1-r_{\eta_c}^2)}} \pi  M^4 f_B C_f f_D
\int_0^1dx_2\int_0^{\infty}b_1b_2db_1db_2\exp(-\frac{\omega_B^2 b_1^2}{2})\nonumber\\&&\{
[\psi ^s(x_2,b_2) \left(r_b-2 x_2\right) r_{\eta _c}+\psi ^v(x_2,b_2) \left(2 r_b-x_2\right)
   \left(r_{\eta _c}^2-1\right)]\nonumber\\&&E_{ab}(t_a)h(\alpha_e,\beta_a,b_1,b_2)S_t(x_2)]
+[\psi ^v(x_2,b_2)  \left(r_{\eta _c}^2+r_c\right)-2 \psi ^s(x_2,b_2)  r_{\eta _c}]\nonumber\\&&
E_{ab}(t_b)h(\alpha_e,\beta_b,b_2,b_1)S_t(x_1)]\},
 \end{eqnarray}
 \begin{eqnarray}
\mathcal {M}_f^{LL}&=&\frac{8\pi  M^4 f_B C_f}{3\sqrt{1-r_{\eta_c}^2}}
\int_0^1dx_2dx_3\int_0^{\infty}b_1b_3db_1db_3\phi _D(x_3,b_3)\exp(-\omega_B^2\frac{b_1^2}{2})\nonumber\\&&
\{[\psi ^s(x_2,b_1) r_{\eta _c} \left(r_c+x_2-1\right)+\psi ^v (x_2,b_1)\left(x_3-x_1-2\left(x_2+x_3-1\right) r_{\eta _c}^2\right)]\nonumber\\&&E_{cd}(t_c)h(\beta_c,\alpha_e,b_3,b_1)
+[-\psi ^s(x_2,b_1) r_{\eta _c} \left(r_c+x_2-1\right)+\psi ^v(x_2,b_1)\nonumber\\&& \left(-2\left( x_3-1\right) r_{\eta
   _c}^2+2r_c+x_2+x_3-2\right)]E_{cd}(t_d)h(\beta_d,\alpha_e,b_3,b_1)
\},
\end{eqnarray}
\begin{eqnarray}
\mathcal {F}_a^{LL}&=&8\pi  M^4 f_B C_f
\int_0^1dx_2dx_3\int_0^{\infty}b_2b_3db_2db_3\phi _D(x_3,b_3)\nonumber\\&&
\{[\psi ^v(x_2,b_2) \left(\left(2 x_3-1\right) r_{\eta _c}^2-x_3+1\right)-2 r_{\eta _c}(r_c+(x_3-2)r_D) \psi ^s(x_2,b_2)
]\nonumber\\&&E_{ef}(t_e)h(\alpha_a,\beta_e,b_2,b_3)
-[2 \left(x_2+1\right) r_D \psi ^s(x_2,b_2) r_{\eta _c}-x_2 \psi ^v (x_2,b_2) \left(r_{\eta
   _c}^2-1\right)]\nonumber\\&&E_{ef}(t_f)h(\alpha_a,\beta_f,b_3,b_2)
\},
\end{eqnarray}
\begin{eqnarray}
\mathcal {M}_a^{LL}&=&\frac{8}{3}\pi  M^4 f_B C_f
\int_0^1dx_2dx_3\int_0^{\infty}b_1b_2db_1db_2\phi _D(x_3,b_2)\exp(-\omega_B^2\frac{b_1^2}{2})\nonumber\\&&
\{[x_2\psi ^v(x_2,b_2)+r_Dr_{\eta_c}(x_2-x_3+1)\psi ^s(x_2,b_2)]E_{gh}(t_g)h(\beta_g,\alpha_a,b_1,b_2)\nonumber\\&&
+[(r_b \left(r_{\eta _c}^2-1\right)-2\left(x_2+ x_3-1\right) r_{\eta _c}^2-x_1+x_3)\psi ^v(x_2,b_2)\nonumber\\&&-\left(4r_b+x_2-x_3-1\right) r_D  r_{\eta _c}\psi ^s(x_2,b_2)]E_{gh}(t_h)h(\beta_h,\alpha_a,b_1,b_2)
\},
\end{eqnarray}
\begin{eqnarray}
\mathcal {F}^{LL}_s&=&\mathcal {F}^{LR}_s=-2 \sqrt{\frac{2}{3}} \pi  M^4 f_B C_f f_{\eta _c}
\int_0^1dx_2\int_0^{\infty}b_1b_2db_1db_2\phi _D (x_2,b_2)\exp(-\frac{\omega_B^2 b_1^2}{2})\nonumber\\&&\{
[\left(2 r_b-2 x_2+1\right) r_{\eta _c}^2+r_D \left(r_b-2 x_2\right)-2
   r_b+x_2]E_{ab}(t_{as})h(\alpha_{es},\beta_{as},b_1,b_2)\nonumber\\&&S_t(x_2)
+(r_c-2r_D)
E_{ab}(t_{bs})h(\alpha_{es},\beta_{bs},b_2,b_1)S_t(x_1)]\},
 \end{eqnarray}
\begin{eqnarray}
\mathcal {M}_s^{LL}&=&\frac{8}{3}\pi  M^4 f_B C_f
\int_0^1dx_2dx_3\int_0^{\infty}b_1b_3db_1db_3\phi _D(x_2,b_1)\psi ^v(x_3,b_3)\exp(-\omega_B^2\frac{b_1^2}{2})\nonumber\\&&
\{[\left(x_2-1\right) r_D+x_3-x_1]E_{cd}(t_{cs})h(\beta_{cs},\alpha_{es},b_3,b_1)\nonumber\\&&
+[-2\left(x_2-1\right) r_{\eta _c}^2+2r_c-\left(x_2-1\right) r_D+x_2+x_3-2]\nonumber\\&&E_{cd}(t_{ds})h(\beta_{ds},\alpha_{es},b_3,b_1)
\},
\end{eqnarray}
\begin{eqnarray}
\mathcal {M}_f^{LR}&=&\frac{8}{3}\pi  M^4 f_B C_f r_D
\int_0^1dx_2dx_3\int_0^{\infty}b_1b_3db_1db_3\phi _D(x_3,b_3)\exp(-\omega_B^2\frac{b_1^2}{2})\times\nonumber\\&&
\{[-\left(x_2-x_3-1\right) \psi ^s(x_2,b_1) r_{\eta _c}+x_3 \psi ^v(x_2,b_1)]E_{cd}(t_c)h(\beta_c,\alpha_e,b_3,b_1)\nonumber\\&&
+[\psi ^s(x_2,b_1)r_{\eta _c} \left(r_c-\left(x_3+x_2-2\right)
   r_D\right)+\psi ^v(x_2,b_1) \left(-r_c+\left(x_3-1\right) r_D\right)]\nonumber\\&&E_{cd}(t_d)h(\beta_d,\alpha_e,b_3,b_1)
\}.
\end{eqnarray}
 \begin{eqnarray}
\mathcal {M}_a^{LR}&=&-\frac{8}{3}\pi  M^4 f_B C_f
\int_0^1dx_2dx_3\int_0^{\infty}b_1b_2db_1db_2\phi _D(x_3,b_2)\exp(-\omega_B^2\frac{b_1^2}{2})\times\nonumber\\&&
\{[\left(x_3-1\right) r_D \psi ^v(x_2,b_2)-\psi ^s (x_2,b_2)r_{\eta _c} \left(2 r_c-x_2\right)]E_{gh}(t_g)h(\beta_g,\alpha_a,b_1,b_2)\nonumber\\&&
-[r_D\left(r_b+x_3\right)\psi ^v (x_2,b_2)-\psi ^s(x_2,b_2) r_{\eta _c}
   \left(r_b-r_c-x_2+1\right)]\nonumber\\&&E_{gh}(t_h)h(\beta_h,\alpha_a,b_1,b_2)
\}.
\end{eqnarray}
\begin{eqnarray}
\mathcal {M}_s^{SP}&=&\frac{8\pi  M^4 f_B C_f}{3\sqrt{1-r_{\eta _c}^2}}
\int_0^1dx_2dx_3\int_0^{\infty}b_1b_3db_1db_3\phi _D(x_2,b_1)\exp(-\omega_B^2\frac{b_1^2}{2})\times\nonumber\\&&
\{\psi ^v(x_3,b_3)[-\left(x_2-1\right) \left(2 r_{\eta _c}^2+r_D-1\right)+2 r_c-x_3]\nonumber\\&&E_{cd}(t_{cs})h(\beta_{cs},\alpha_{es},b_3,b_1)
-[r_c \psi ^s (x_3,b_3)r_{\eta _c}+\psi ^v (x_3,b_3)\nonumber\\&&\left(-\left(x_2-1\right) r_D+x_1+x_3-1\right)]E_{cd}(t_{ds})h(\beta_{ds},\alpha_{es},b_3,b_1)
\},
\end{eqnarray}
\begin{eqnarray}
\mathcal {F}_a^{SP}&=&-\frac{16 \pi  M^4 f_B C_f}{\sqrt{1-r_{\eta _c}^2}}
\int_0^1dx_2dx_3\int_0^{\infty}b_2b_3db_2db_3\phi _D(x_3,b_3)\times\nonumber\\&&
\{[\psi ^v(x_2,b_2) \left(r_c+\left(x_3-1\right) r_D\right)-2 \psi ^s(x_2,b_2) r_{\eta _c}]E_{ef}(t_e)h(\alpha_a,\beta_e,b_2,b_3)\nonumber\\&&
-[x_2 r_{\eta _c}\psi ^s(x_2,b_2)+2r_D\psi ^v(x_2,b_2)]E_{ef}(t_f)h(\alpha_a,\beta_f,b_3,b_2)
\},
\end{eqnarray}
\begin{eqnarray}
\mathcal {F}^{SP}_f&=&-4 \sqrt{\frac{2}{3}} \pi  M^4 f_B C_f f_Dr_D
\int_0^1dx_2\int_0^{\infty}b_1b_2db_1db_2\exp(-\frac{\omega_B^2 b_1^2}{2})\nonumber\\&&\{
[\left(r_b-2\right) \psi ^v(x_2,b_2) -\psi ^s(x_2,b_2)  \left(4 r_b-x_2-1\right) r_{\eta _c}]\nonumber\\&&E_{ab}(t_a)h(\alpha_e,\beta_a,b_1,b_2)S_t(x_2)]
-2 \psi ^s(x_2,b_2)  r_{\eta _c}\nonumber\\&&
E_{ab}(t_b)h(\alpha_e,\beta_b,b_2,b_1)S_t(x_1)]\},
 \end{eqnarray}
where
\begin{eqnarray}\label{eq:betai11}
\alpha_e&=&-[x_1+r_{\eta_c}^2(x_2-1)][x_1+x_2-1+(1-x_2)r_D^2]M_B^2,\nonumber\\ \alpha_{es}&=&-[x_1+r_{D}^2(x_2-1)][x_1+x_2-1+(1-x_2)r_{\eta_c}^2]M_B^2, \nonumber\\
\beta_a&=&[r_b^2-(1+r_{\eta_c}^2(x_2-1))(x_2-r_D^2(x_2-1))]M_B^2, \nonumber\\ \beta_b&=&[r_c^2+(r_{\eta_c}^2-x_1)(r_D^2+x_1-1)]M_B^2,\nonumber\\
\beta_c&=&-[x_1+x_2-1+(1-x_2-x_3)r_D^2][r_{\eta_c}^2(x_2+x_3-1)-x_3+x_1]M_B^2,\nonumber\\\quad \beta_d&=&r_c^2M_B^2-[x_1+x_2-1-(x_2-x_3)r_D^2][x_1+x_3-1+r_{\eta_c}^2(x_2-x_3)]M_B^2,\nonumber\\
\beta_{as}&=&[r_b^2-(1+r_D^2(x_2-1))(x_2-r_{\eta_c}^2(x_2-1))]M_B^2, \nonumber\\ \beta_{bs}&=&[(r_D^2-x_1)(r_{\eta_c}^2+x_1-1)]M_B^2,\nonumber\\
\beta_{cs}&=&r_c^2M_B^2-[x_1+x_2-1+(1-x_2-x_3)r_{\eta_c}^2][r_D^2(x_2+x_3-1)-x_3+x_1]M_B^2,\nonumber\\\quad \beta_{ds}&=&r_c^2M_B^2-[x_1+x_2-1-(x_2-x_3)r_{\eta_c}^2][x_1+x_3-1+r_D^2(x_2-x_3)]M_B^2,\nonumber\\
\alpha_a&=&-[1-x_3+r_{\eta_c}^2(x_2+x_3-1)][x_2-r_D^2(x_2+x_3-1)]M_B^2,\nonumber\\
\beta_e&=&[r_c^2-(1+(r_{\eta_c}^2-1)x_3)(1-r_D^2x_3)]M_B^2,\nonumber\\ \beta_f&=&[1+r_{\eta_c}^2(x_2-1)][x_2-r_D^2(x_2-1)]M_B^2,\nonumber\\
\beta_g&=&r_c^2M_B^2-(r_{\eta_c}^2(1-x_3-x_2)+x_1+x_3-1)(r_D^2(x_2+x_3-1)+x_1-x_2)M_B^2,\nonumber\\
\beta_h&=&r_b^2M_B^2-(r_{\eta_c}^2(x_2+x_3-1)-x_3+x_1)(r_D^2(1-x_2-x_3)+x_1+x_2-1)M_B^2.\nonumber\\&&
\end{eqnarray}

\subsection{Factorization formulas for $B_c\rightarrow \eta_c D^*_{(s)}$}\label{sec:apD}
\begin{eqnarray}
\mathcal {F}^{LL}_f&=&2 \sqrt{\frac{2}{3(1-r_{\eta_c}^2)}} \pi  M^4 f_B C_f f_D
\int_0^1dx_2\int_0^{\infty}b_1b_2db_1db_2\exp(-\frac{\omega_B^2 b_1^2}{2})\nonumber\\&&\{
[\psi ^s(x_2,b_2) \left(r_b-2 x_2\right) r_{\eta _c}+\psi ^v(x_2,b_2) \left(2 r_b-x_2\right)
   \left(r_{\eta _c}^2-1\right)]\nonumber\\&&E_{ab}(t_a)h(\alpha_e,\beta_a,b_1,b_2)S_t(x_2)]
+[\psi ^v(x_2,b_2)  \left(r_{\eta _c}^2+r_c\right)-2 \psi ^s(x_2,b_2)  r_{\eta _c}]\nonumber\\&&
E_{ab}(t_b)h(\alpha_e,\beta_b,b_2,b_1)S_t(x_1)]\},
 \end{eqnarray}
 \begin{eqnarray}
\mathcal {M}_f^{LL}&=&\frac{8\pi  M^4 f_B C_f}{3\sqrt{1-r_{\eta_c}^2}}
\int_0^1dx_2dx_3\int_0^{\infty}b_1b_3db_1db_3\phi _D(x_3,b_3)\exp(-\omega_B^2\frac{b_1^2}{2})\nonumber\\&&
\{[\psi ^s(x_2,b_1) r_{\eta _c} \left(r_c+x_2-1\right)+\psi ^v (x_2,b_1)\left(x_3-x_1-2\left(x_2+x_3-1\right) r_{\eta _c}^2\right)]\nonumber\\&&E_{cd}(t_c)h(\beta_c,\alpha_e,b_3,b_1)
+[-\psi ^s(x_2,b_1) r_{\eta _c} \left(r_c+x_2-1\right)+\psi ^v(x_2,b_1)\nonumber\\&& \left(-2\left(x_2+ x_3-1\right) r_{\eta
   _c}^2+2r_c+x_2+x_3-2\right)]E_{cd}(t_d)h(\beta_d,\alpha_e,b_3,b_1)
\},
\end{eqnarray}
\begin{eqnarray}
\mathcal {F}_a^{LL}&=&\frac{8\pi  M^4 f_B C_f}{\sqrt{1-r_{\eta _c}^2}}
\int_0^1dx_2dx_3\int_0^{\infty}b_2b_3db_2db_3\phi _D(x_3,b_3)\nonumber\\&&
\{[\psi ^v(x_2,b_2) \left(\left(2 x_3-1\right) r_{\eta _c}^2-x_3+1\right)-2 r_{\eta _c}(r_c+x_3r_D) \psi ^s(x_2,b_2)
]E_{ef}(t_e)\nonumber\\&&h(\alpha_a,\beta_e,b_2,b_3)
-x_2\psi ^v(x_2,b_2) (1-r_{\eta _c}^2)E_{ef}(t_f)h(\alpha_a,\beta_f,b_3,b_2)
\},
\end{eqnarray}
\begin{eqnarray}
\mathcal {M}_a^{LL}&=&\frac{8\pi  M^4 f_B C_f}{3\sqrt{1-r_{\eta _c}^2}}
\int_0^1dx_2dx_3\int_0^{\infty}b_1b_2db_1db_2\phi _D(x_3,b_2)\exp(-\omega_B^2\frac{b_1^2}{2})\nonumber\\&&
\{[x_2\psi ^v(x_2,b_2)+r_Dr_{\eta_c}(x_2+x_3-1)\psi ^s(x_2,b_2)]E_{gh}(t_g)h(\beta_g,\alpha_a,b_1,b_2)\nonumber\\&&
-[(r_b \left(r_{\eta _c}^2-1\right)-2\left(x_2+ x_3-1\right) r_{\eta _c}^2+x_3-x_1)\psi ^v(x_2,b_2)\nonumber\\&&+\left(x_2+x_3-1\right) r_D  r_{\eta _c}\psi ^s(x_2,b_2)]E_{gh}(t_h)h(\beta_h,\alpha_a,b_1,b_2)
\},
\end{eqnarray}
\begin{eqnarray}
\mathcal {F}^{LL}_s=\mathcal {F}^{LR}_s&=&-2 \sqrt{\frac{2}{3(1-r_{\eta _c}^2)}} \pi  M^4 f_B C_f f_{\eta _c}
\int_0^1dx_2\int_0^{\infty}b_1b_2db_1db_2\phi _D (x_2,b_2)\exp(-\frac{\omega_B^2 b_1^2}{2})\nonumber\\&&\{
[\left(2 r_b-2 x_2+1\right) r_{\eta _c}^2+r_D \left(r_b-2 x_2\right)-2
   r_b+x_2]\nonumber\\&&E_{ab}(t_{as})h(\alpha_{es},\beta_{as},b_1,b_2)S_t(x_2)
-r_c
E_{ab}(t_{bs})h(\alpha_{es},\beta_{bs},b_2,b_1)S_t(x_1)]\},
 \end{eqnarray}
\begin{eqnarray}
\mathcal {M}_s^{LL}&=&-\frac{8\pi  M^4 f_B C_f}{3\sqrt{1-r_{\eta _c}^2}}
\int_0^1dx_2dx_3\int_0^{\infty}b_1b_3db_1db_3\phi _D(x_2,b_1)\psi ^v(x_3,b_3)\exp(-\omega_B^2\frac{b_1^2}{2})\nonumber\\&&
\{[\left(x_2-1\right) r_D-x_3+x_1]E_{cd}(t_{cs})h(\beta_{cs},\alpha_{es},b_3,b_1)\nonumber\\&&
-[-2\left(x_2-1\right) r_{\eta _c}^2+2r_c-\left(x_2-1\right) r_D+x_2+x_3-2]\nonumber\\&&E_{cd}(t_{ds})h(\beta_{ds},\alpha_{es},b_3,b_1)
\},
\end{eqnarray}
\begin{eqnarray}
\mathcal {M}_f^{LR}&=&\frac{8\pi  M^4 f_B C_f r_D}{3\sqrt{1-r_{\eta _c}^2}}
\int_0^1dx_2dx_3\int_0^{\infty}b_1b_3db_1db_3\phi _D(x_3,b_3)\exp(-\omega_B^2\frac{b_1^2}{2})\times\nonumber\\&&
\{[\left(x_2+x_3-1\right) \psi ^s(x_2,b_1) r_{\eta _c}+x_3 \psi ^v(x_2,b_1)]E_{cd}(t_c)h(\beta_c,\alpha_e,b_3,b_1)\nonumber\\&&
-[\psi ^s(x_2,b_1)r_{\eta _c} \left(r_c+\left(x_3-x_2\right)
   r_D\right)+\psi ^v(x_2,b_1) \left(r_c+\left(x_3-1\right) r_D\right)]\nonumber\\&&E_{cd}(t_d)h(\beta_d,\alpha_e,b_3,b_1)
\}.
\end{eqnarray}
 \begin{eqnarray}
\mathcal {M}_a^{LR}&=&-\frac{8\pi  M^4 f_B C_f}{3\sqrt{1-r_{\eta _c}^2}}
\int_0^1dx_2dx_3\int_0^{\infty}b_1b_2db_1db_2\phi _D(x_3,b_2)\exp(-\omega_B^2\frac{b_1^2}{2})\times\nonumber\\&&
\{[\left(x_3-1\right) r_D \psi ^v(x_2,b_2)-\psi ^s (x_2,b_2)r_{\eta _c} \left(2 r_c-x_2\right)]E_{gh}(t_g)h(\beta_g,\alpha_a,b_1,b_2)\nonumber\\&&
-[r_D\left(r_b+x_3\right)\psi ^v (x_2,b_2)-\psi ^s(x_2,b_2) r_{\eta _c}
   \left(r_b-r_c-x_2+1\right)]\nonumber\\&&E_{gh}(t_h)h(\beta_h,\alpha_a,b_1,b_2)
\}.
\end{eqnarray}
\begin{eqnarray}
\mathcal {M}_s^{SP}&=&\frac{8\pi  M^4 f_B C_f}{3\sqrt{1-r_{\eta _c}^2}}
\int_0^1dx_2dx_3\int_0^{\infty}b_1b_3db_1db_3\phi _D(x_2,b_1)\exp(-\omega_B^2\frac{b_1^2}{2})\times\nonumber\\&&
\{\psi ^v(x_3,b_3)[-\left(x_2-1\right) \left(2 r_{\eta _c}^2+r_D-1\right)+2
   r_c-x_3]\nonumber\\&&E_{cd}(t_{cs})h(\beta_{cs},\alpha_{es},b_3,b_1)
-[r_c \psi ^s (x_3,b_3)r_{\eta _c}+\psi ^v (x_3,b_3)\nonumber\\&&\left(\left(x_2-1\right) r_D+x_1+x_3-1\right)]E_{cd}(t_{ds})h(\beta_{ds},\alpha_{es},b_3,b_1)
\},
\end{eqnarray}
\begin{eqnarray}
\mathcal {F}_a^{SP}&=&-\frac{16 \pi  M^4 f_B C_f}{\sqrt{1-r_{\eta _c}^2}}
\int_0^1dx_2dx_3\int_0^{\infty}b_2b_3db_2db_3\phi _D(x_3,b_3)\times\nonumber\\&&
\{[\psi ^v(x_2,b_2) \left(r_c-\left(x_3-1\right) r_D\right)-2 \psi ^s(x_2,b_2) r_{\eta _c}]\nonumber\\&&E_{ef}(t_e)h(\alpha_a,\beta_e,b_2,b_3)
-x_2 r_{\eta _c}\psi ^s(x_2,b_2)E_{ef}(t_f)h(\alpha_a,\beta_f,b_3,b_2)
\},
\end{eqnarray}
where the expressions of  $\beta$ and $\alpha$ are the same as those of Eq. (\ref{eq:betai11}).
\subsection{Factorization formulas for $B_c\rightarrow J/\Psi D_{(s)}$}\label{sec:avD}
\begin{eqnarray}
\mathcal {F}^{LL}_f&=&2 \sqrt{\frac{2}{3}} \pi  M^4 f_B C_f f_D
\int_0^1dx_2\int_0^{\infty}b_1b_2db_1db_2\exp(-\frac{\omega_B^2 b_1^2}{2})\nonumber\\&&\{
[\psi ^L (x_2,b_2)\left(r_{J/\psi }^2-1\right) \left(2 r_b-x_2\right)+r_{J/\psi }
   \psi ^t(x_2,b_2) \left(r_b-2 x_2\right)]\nonumber\\&&E_{ab}(t_a)h(\alpha_e,\beta_a,b_1,b_2)S_t(x_2)]
\nonumber\\&&-[\psi ^L(x_2,b_2) \left(r_c+r_{\psi }^2\right) ]
E_{ab}(t_b)h(\alpha_e,\beta_b,b_2,b_1)S_t(x_1)]\},
 \end{eqnarray}
\begin{eqnarray}
\mathcal {M}_f^{LL}&=&\frac{8}{3}\pi  M^4 f_B C_f
\int_0^1dx_2dx_3\int_0^{\infty}b_1b_3db_1db_3\phi _D(x_3,b_3)\exp(-\omega_B^2\frac{b_1^2}{2})\times\nonumber\\&&
\{[r_{J/\psi } \psi ^t(x_2,b_1)
   \left(r_c+x_2-1\right)-\psi ^L (x_2,b_1)\left(x_3 \left(1-2 r_{\psi }^2\right)-r_c\right)]\nonumber\\&&E_{cd}(t_c)h(\beta_c,\alpha_e,b_3,b_1)
+[r_{J/\psi } \psi ^t(x_2,b_1)
   \left(r_c+x_2-1\right)-\nonumber\\&&\psi ^L(x_2,b_1) \left(2 r_c-2 \left(x_3-1\right) r_{\psi }^2+x_2+x_3-2\right)]\nonumber\\&&E_{cd}(t_d)h(\beta_d,\alpha_e,b_3,b_1)
\},
\end{eqnarray}
\begin{eqnarray}
\mathcal {F}_a^{LL}&=&-8\pi  M^4 f_B C_f
\int_0^1dx_2dx_3\int_0^{\infty}b_2b_3db_2db_3\phi _D(x_3,b_3)\times\nonumber\\&&
\{[\psi ^L (x_2,b_2)\left(\left(2 x_3-1\right) r_{J/\psi }^2-x_3+1\right)]E_{ef}(t_e)h(\alpha_a,\beta_e,b_2,b_3)\nonumber\\&&
-[2 \left(x_2-1\right) r_D r_{J/\psi } \psi
   ^t(x_2,b_2)-x_2 \psi ^L(x_2,b_2) \left(r_{J/\psi }^2-1\right)]\nonumber\\&&E_{ef}(t_f)h(\alpha_a,\beta_f,b_3,b_2)
\},
\end{eqnarray}
\begin{eqnarray}
\mathcal {M}_a^{LL}&=&\frac{8}{3} \pi  M^4 f_B C_f
\int_0^1dx_2dx_3\int_0^{\infty}b_1b_2db_1db_2\phi _D(x_3,b_2)\exp(-\omega_B^2\frac{b_1^2}{2})\times\nonumber\\&&
\{[x_2 \psi ^L(x_2,b_2) \left(2 r_{J/\psi
   }^2-1\right)-\left(x_2+x_3-1\right) r_D r_{J/\psi } \psi ^t(x_2,b_2)]\nonumber\\&&E_{gh}(t_g)h(\beta_g,\alpha_a,b_1,b_2)
-[\psi ^L (x_2,b_2)(r_b \left(r_{\psi }^2-1\right)-r_c+x_3 \left(1-2 r_{\psi
   }^2\right))\nonumber\\&&+\left(x_2+x_3-1\right) r_D r_{J/\psi } \psi ^t(x_2,b_2)]E_{gh}(t_h)h(\beta_h,\alpha_a,b_1,b_2)
\},
\end{eqnarray}
\begin{eqnarray}
\mathcal {F}^{LL}_s=\mathcal {F}^{LR}_s&=&2 \sqrt{\frac{2}{3}} \pi  M^4 f_B C_f f_{J/\psi}
\int_0^1dx_2\int_0^{\infty}b_1b_2db_1db_2\phi _D (x_2,b_2)\exp(-\frac{\omega_B^2 b_1^2}{2})\nonumber\\&&\{
[r_D \left(r_b-2 x_2\right)+r_{\psi }^2 \left(2 r_b-2 x_2+1\right)-2 r_b+x_2]\nonumber\\&&E_{ab}(t_{as})h(\alpha_{es},\beta_{as},b_1,b_2)S_t(x_2)
+\nonumber\\&&(r_c-2r_D)
E_{ab}(t_{bs})h(\alpha_{es},\beta_{bs},b_2,b_1)S_t(x_1)]\},
 \end{eqnarray}
\begin{eqnarray}
\mathcal {M}_s^{LL}&=&-\frac{8}{3}\pi  M^4 f_B C_f
\int_0^1dx_2dx_3\int_0^{\infty}b_1b_3db_1db_3\phi _D(x_2,b_1)\psi ^L(x_3,b_3)\exp(-\omega_B^2\frac{b_1^2}{2})\nonumber\\&&
\{[ -x_1+\left(x_2-1\right)
   r_D+x_3 \left(1-2 r_{J/\psi }^2\right)]E_{cd}(t_{cs})h(\beta_{cs},\alpha_{es},b_3,b_1)\nonumber\\&&
+[2r_c-\left(x_2-1\right)
   r_D-2\left( x_2-1\right) r_{J/\psi }^2+x_2+x_3-2]\nonumber\\&&E_{cd}(t_{ds})h(\beta_{ds},\alpha_{es},b_3,b_1)
\},
\end{eqnarray}
\begin{eqnarray}
\mathcal {M}_f^{LR}&=&\frac{8 }{3}\pi  M^4 f_B C_f
\int_0^1dx_2dx_3\int_0^{\infty}b_1b_3db_1db_3\phi _D(x_3,b_3)\exp(-\omega_B^2\frac{b_1^2}{2})\times\nonumber\\&&
\{r_D[x_3 \psi
   ^L(x_2,b_1)-\left(x_2+x_3-1\right) r_{J/\psi } \psi ^t(x_2,b_1)]E_{cd}(t_c)h(\beta_c,\alpha_e,b_3,b_1)
\nonumber\\&&+[\psi ^L(x_2,b_1) \left(\left(x_3-1\right)
   r_D-r_c\right)+r_{J/\psi } \psi ^t(x_2,b_1) \left(r_c+\left(x_2-x_3\right)
   r_D\right)]\nonumber\\&&E_{cd}(t_d)h(\beta_d,\alpha_e,b_3,b_1)
\},
\end{eqnarray}
\begin{eqnarray}
\mathcal {M}_a^{LR}&=&-\frac{8}{3}\pi  M^4 f_B C_f
\int_0^1dx_2dx_3\int_0^{\infty}b_1b_2db_1db_2\phi _D(x_3,b_2)\exp(-\omega_B^2\frac{b_1^2}{2})\times\nonumber\\&&
\{[\left(x_3-1\right) r_D \psi
   ^L(x_2,b_2)-r_{J/\psi } \psi ^t(x_2,b_2) \left(2 r_c-x_2\right)]E_{gh}(t_g)h(\beta_g,\alpha_a,b_1,b_2)\nonumber\\&&
-[r_D \psi ^L(x_2,b_2)
   \left(r_b+x_3\right)-r_{J/\psi } \psi ^t(x_2,b_2) \left(r_b-r_c-x_2+1\right)]\nonumber\\&&E_{gh}(t_h)h(\beta_h,\alpha_a,b_1,b_2)
\},
\end{eqnarray}
\begin{eqnarray}
\mathcal {M}_s^{SP}&=&\frac{8}{3}\pi  M^4 f_B C_f
\int_0^1dx_2dx_3\int_0^{\infty}b_1b_3db_1db_3\phi _D(x_2,b_1)\exp(-\omega_B^2\frac{b_1^2}{2})\nonumber\\&&
\{[2 r_c-\left(x_2-1\right) \left(r_D+2 r_{\psi }^2-1\right)-x_3]\psi ^L(x_3,b_3)\nonumber\\&&E_{cd}(t_{cs})h(\beta_{cs},\alpha_{es},b_3,b_1)
+[r_c r_{J/\psi } \psi ^t(x_3,b_3)+\psi ^L(x_3,b_3)\nonumber\\&&
   \left(-r_c+\left(x_2-1\right) r_D+\left(x_3-1\right) \left(2
   r_{\psi }^2-1\right)\right)]\nonumber\\&&E_{cd}(t_{ds})h(\beta_{ds},\alpha_{es},b_3,b_1)
\},
\end{eqnarray}
\begin{eqnarray}
\mathcal {F}_a^{SP}&=&16\pi  M^4 f_B C_f
\int_0^1dx_2dx_3\int_0^{\infty}b_2b_3db_2db_3\phi _D(x_3,b_3)\times\nonumber\\&&
\{\psi ^L (x_2,b_2)\left(r_c-\left(x_3-1\right) r_D\right)E_{ef}(t_e)h(\alpha_a,\beta_e,b_2,b_3)\nonumber\\&&
+[2 r_D \psi ^L(x_2,b_2)-x_2 r_{J/\psi } \psi ^t(x_2,b_2)]E_{ef}(t_f)h(\alpha_a,\beta_f,b_3,b_2)
\},
\end{eqnarray}
\begin{eqnarray}
\mathcal {F}^{SP}_f&=&-4 \sqrt{\frac{2}{3}} \pi  M^4 f_B C_f f_D r_D
\int_0^1dx_2\int_0^{\infty}b_1b_2db_1db_2\exp(-\frac{\omega_B^2 b_1^2}{2})\nonumber\\&&
[\left(r_b-2\right)
   \psi ^L(x_2,b_2)-\left(x_2-1\right) r_{J/\psi } \psi ^t(x_2,b_2)]\nonumber\\&&E_{ab}(t_a)h(\alpha_e,\beta_a,b_1,b_2)S_t(x_2),
 \end{eqnarray}
where the expressions of  $\beta_{a,b,c,d}$ and $\alpha_e$ are the similar to those of Eq. (\ref{eq:betai11}),
 but with the replacement $r_{\eta_c}\rightarrow r_{J/\psi}$.

\subsection{Factorization formulas for $B_c\rightarrow J/\Psi D^*_{(s)}$}\label{sec:avDstar}
\begin{eqnarray}
\mathcal {F}^{LL,L}_f&=&2 \sqrt{\frac{2}{3(1-r_{J/\psi}^2)}} \pi  M^4 f_B C_f f_D
\int_0^1dx_2\int_0^{\infty}b_1b_2db_1db_2\exp(-\frac{\omega_B^2 b_1^2}{2})\nonumber\\&&\{
[\psi ^L (x_2,b_2)\left(r_{J/\psi }^2-1\right) \left(2 r_b-x_2\right)+r_{J/\psi }
   \psi ^t(x_2,b_2) \left(r_b-2 x_2\right)]\nonumber\\&&E_{ab}(t_a)h(\alpha_e,\beta_a,b_1,b_2)S_t(x_2)]
\nonumber\\&&-[\psi ^L(x_2,b_2) \left(r_c+r_{\psi }^2\right) ]
E_{ab}(t_b)h(\alpha_e,\beta_b,b_2,b_1)S_t(x_1)]\},
 \end{eqnarray}
\begin{eqnarray}
\mathcal {M}_f^{LL,L}&=&\frac{8\pi  M^4 f_B C_f}{3\sqrt{(1-r_{J/\psi}^2)}}
\int_0^1dx_2dx_3\int_0^{\infty}b_1b_3db_1db_3\phi _D(x_3,b_3)\exp(-\omega_B^2\frac{b_1^2}{2})\times\nonumber\\&&
\{[r_{J/\psi }
   \psi ^t(x_2,b_1) \left(r_c+x_2-1\right)-\psi ^L(x_2,b_1) \left(r_c+x_3 \left(2 r_{\psi }^2-1\right)\right)]\nonumber\\&&E_{cd}(t_c)h(\beta_c,\alpha_e,b_3,b_1)
-[\psi ^L(x_2,b_1) \left(2r_c-2\left( x_3-1\right) r_{J/\psi
   }^2+x_2+x_3-2\right)\nonumber\\&&-r_{J/\psi } \psi ^t (x_2,b_1)\left(r_c+x_2-1\right)]E_{cd}(t_d)h(\beta_d,\alpha_e,b_3,b_1)
\},
\end{eqnarray}
\begin{eqnarray}
\mathcal {F}_a^{LL,L}&=&-\frac{8\pi  M^4 f_B C_f}{\sqrt{1-r_{J/\psi}^2}}
\int_0^1dx_2dx_3\int_0^{\infty}b_2b_3db_2db_3\phi _D(x_3,b_3)\psi ^L (x_2,b_2)\nonumber\\&&
\{[\left(\left(2 x_3-1\right) r_{J/\psi }^2-x_3+1\right)]E_{ef}(t_e)h(\alpha_a,\beta_e,b_2,b_3)\nonumber\\&&
- x_2\left(1-r_{J/\psi }^2\right)E_{ef}(t_f)h(\alpha_a,\beta_f,b_3,b_2)
\},
\end{eqnarray}
\begin{eqnarray}
\mathcal {M}_a^{LL,L}&=&\frac{8\pi  M^4 f_B C_f}{3\sqrt{1-r_{J/\psi}^2}}
\int_0^1dx_2dx_3\int_0^{\infty}b_1b_2db_1db_2\phi _D(x_3,b_2)\exp(-\omega_B^2\frac{b_1^2}{2})\times\nonumber\\&&
\{[x_2 \psi ^L(x_2,b_2) \left(2
   r_{J/\psi }^2-1\right)-\left(x_2-x_3+1\right) r_D r_{J/\psi } \psi ^t(x_2,b_2)]\nonumber\\&&E_{gh}(t_g)h(\beta_g,\alpha_a,b_1,b_2)
+[\left(x_2-x_3-1\right) r_D r_{J/\psi } \psi
   ^t(x_2,b_2)-\psi ^L(x_2,b_2)\nonumber\\&& \left(r_b \left(r_{J/\psi }^2-1\right)-r_c-x_3 \left(2 r_{\psi }^2-1\right)\right)]E_{gh}(t_h)h(\beta_h,\alpha_a,b_1,b_2)
\},
\end{eqnarray}
\begin{eqnarray}
\mathcal {F}^{LL,L}_s&=&\mathcal {F}^{LR,L}_s=2 \sqrt{\frac{2}{3(1-r_{J/\psi}^2)}} \pi  M^4 f_B C_f f_{J/\psi}
\int_0^1dx_2\int_0^{\infty}b_1b_2db_1db_2\phi _D (x_2,b_2)\nonumber\\&&\exp(-\frac{\omega_B^2 b_1^2}{2})\{
[r_D \left(r_b-2 x_2\right)+r_{J/\psi }^2 \left(2 r_b-2 x_2+1\right)-2 r_b+x_2]\nonumber\\&&E_{ab}(t_{as})h(\alpha_{es},\beta_{as},b_1,b_2)S_t(x_2)
-r_c
E_{ab}(t_{bs})h(\alpha_{es},\beta_{bs},b_2,b_1)S_t(x_1)]\},
 \end{eqnarray}
\begin{eqnarray}
\mathcal {M}_s^{LL,L}&=&\frac{8\pi  M^4 f_B C_f}{3\sqrt{1-r_{J/\psi}^2}}
\int_0^1dx_2dx_3\int_0^{\infty}b_1b_3db_1db_3\phi _D(x_2,b_1)\psi ^L(x_3,b_3)\exp(-\omega_B^2\frac{b_1^2}{2})\nonumber\\&&
\{[x_1+\left(x_2-1\right)
   r_D+ x_3 \left(2 r_{J/\psi }^2-1\right)]E_{cd}(t_{cs})h(\beta_{cs},\alpha_{es},b_3,b_1)\nonumber\\&&
-[2r_c-\left(x_2-1\right) r_D-2\left( x_2-1\right) r_{J/\psi
   }^2+x_2+x_3-2]\nonumber\\&&E_{cd}(t_{ds})h(\beta_{ds},\alpha_{es},b_3,b_1)
\},
\end{eqnarray}
\begin{eqnarray}
\mathcal {M}_f^{LR,L}&=&\frac{8\pi  M^4 f_B C_f }{3\sqrt{1-r_{J/\psi}^2}}
\int_0^1dx_2dx_3\int_0^{\infty}b_1b_3db_1db_3\phi _D(x_3,b_3)\exp(-\omega_B^2\frac{b_1^2}{2})\times\nonumber\\&&
\{[r_D\left(x_3 \psi ^L(x_2,b_1)+\left(x_2-x_3-1\right) r_{J/\psi } \psi ^t(x_2,b_1)\right)]E_{cd}(t_c)h(\beta_c,\alpha_e,b_3,b_1)
\nonumber\\&&+[r_{J/\psi } \psi ^t(x_2,b_1) \left(r_c+\left(x_2+x_3-2\right) r_D\right)-\psi ^L(x_2,b_1)
   \left(r_c+\left(x_3-1\right) r_D\right)]\nonumber\\&&E_{cd}(t_d)h(\beta_d,\alpha_e,b_3,b_1)
\},
\end{eqnarray}
 \begin{eqnarray}
\mathcal {M}_a^{LR,L}&=&-\frac{8\pi  M^4 f_B C_f}{3\sqrt{1-r_{J/\psi}^2}}
\int_0^1dx_2dx_3\int_0^{\infty}b_1b_2db_1db_2\phi _D(x_3,b_2)\exp(-\omega_B^2\frac{b_1^2}{2})\times\nonumber\\&&
\{[\left(x_3-1\right) r_D \psi
   ^L(x_2,b_2)-r_{J/\psi } \psi ^t (x_2,b_2)\left(2 r_c -x_2\right)]E_{gh}(t_g)h(\beta_g,\alpha_a,b_1,b_2)\nonumber\\&&
-[r_{J/\psi } \psi ^t(x_2,b_2) \left(-r_b+r_c+x_2-1\right)+r_D \psi ^L(x_2,b_2)
   \left(r_b+x_3\right)]\nonumber\\&&E_{gh}(t_h)h(\beta_h,\alpha_a,b_1,b_2)
\},
\end{eqnarray}
\begin{eqnarray}
\mathcal {M}_s^{SP,L}&=&\frac{8\pi  M^4 f_B C_f}{3\sqrt{1-r_{J/\psi}^2}}
\int_0^1dx_2dx_3\int_0^{\infty}b_1b_3db_1db_3\phi _D(x_2,b_1)\exp(-\omega_B^2\frac{b_1^2}{2})\times\nonumber\\&&
\{[\psi ^L (x_3,b_3)\left(2r_c-\left(x_2-1\right) \left(r_D+2 r_{J/\psi
   }^2-1\right)-x_3 \right)]\nonumber\\&&E_{cd}(t_{cs})h(\beta_{cs},\alpha_{es},b_3,b_1)
+[r_c r_{J/\psi } \psi ^t(x_3,b_3)+\psi ^L (x_3,b_3)\nonumber\\&&\left(\left(x_3-1\right) \left(2
   r_{J/\psi }^2-1\right)-\left(x_2-1\right) r_D-r_c(1-r_{}J/\psi)\right)]\nonumber\\&&E_{cd}(t_{ds})h(\beta_{ds},\alpha_{es},b_3,b_1)
\},
\end{eqnarray}
\begin{eqnarray}
\mathcal {F}_a^{SP,L}&=&\frac{16 \pi  M^4 f_B C_f}{\sqrt{1-r_{J/\psi}^2}}
\int_0^1dx_2dx_3\int_0^{\infty}b_2b_3db_2db_3\phi _D(x_3,b_3)\times\nonumber\\&&
\{[\psi ^L (x_2,b_2)\left(r_c+\left(x_3-1\right) r_D\right)]E_{ef}(t_e)h(\alpha_a,\beta_e,b_2,b_3)
-\nonumber\\&&[x_2 r_{J/\psi } \psi ^t(x_2,b_2)]E_{ef}(t_f)h(\alpha_a,\beta_f,b_3,b_2)
\},\nonumber\\
\end{eqnarray}
\begin{eqnarray}
\mathcal {F}^{LL,N}_f&=&2 \sqrt{\frac{2}{3}} \pi  M^4 f_B C_f f_Dr_D
\int_0^1dx_2\int_0^{\infty}b_1b_2db_1db_2\exp(-\frac{\omega_B^2 b_1^2}{2})\nonumber\\&&\{
[\left(r_{\psi } \psi ^T
   \left(-4 r_b+x_2+1\right)+\left(r_b-2\right) \psi ^V\right)]\nonumber\\&&E_{ab}(t_a)h(\alpha_e,\beta_a,b_1,b_2)S_t(x_2)]
\nonumber\\&&-[ r_{\psi } \psi ^T ]
E_{ab}(t_b)h(\alpha_e,\beta_b,b_2,b_1)S_t(x_1)]\},
 \end{eqnarray}
\begin{eqnarray}
\mathcal {F}^{LL,T}_f&=&2 \sqrt{\frac{2}{3}} \pi  M^4 f_B C_f f_Dr_D
\int_0^1dx_2\int_0^{\infty}b_1b_2db_1db_2\exp(-\frac{\omega_B^2 b_1^2}{2})\nonumber\\&&\{
[\left(\left(r_b-2\right) \psi ^V-\left(x_2-1\right) r_{\psi } \psi
   ^T\right)]\nonumber\\&&E_{ab}(t_a)h(\alpha_e,\beta_a,b_1,b_2)S_t(x_2)]
\nonumber\\&&-[ r_{\psi } \psi ^T ]
E_{ab}(t_b)h(\alpha_e,\beta_b,b_2,b_1)S_t(x_1)]\},
 \end{eqnarray}
\begin{eqnarray}
\mathcal {M}_f^{LL,N}&=&\frac{8}{3} \pi  M^4 f_B C_f r_D
\int_0^1dx_2dx_3\int_0^{\infty}b_1b_3db_1db_3\phi _D(x_3,b_3)\exp(-\omega_B^2\frac{b_1^2}{2})\times\nonumber\\&&
\{[(x_3\psi ^V(x_2,b_1)]E_{cd}(t_c)h(\beta_c,\alpha_e,b_3,b_1)\nonumber\\&&
+[\left(2
   \left(x_2+x_3-2\right) r_{\psi } \psi ^T-\left(x_3-1\right) \psi
   ^V\right)]E_{cd}(t_d)h(\beta_d,\alpha_e,b_3,b_1)
\}.
\end{eqnarray}
\begin{eqnarray}
\mathcal {M}_f^{LL,T}&=&\frac{8}{3} \pi  M^4 f_B C_f r_D
\int_0^1dx_2dx_3\int_0^{\infty}b_1b_3db_1db_3\phi _D(x_3,b_3)\exp(-\omega_B^2\frac{b_1^2}{2})\times\nonumber\\&&
\{(x_3\psi ^V(x_2,b_1))E_{cd}(t_c)h(\beta_c,\alpha_e,b_3,b_1)\nonumber\\&&
+[\psi ^V(x_2,b_1) (1-x_3)]E_{cd}(t_d)h(\beta_d,\alpha_e,b_3,b_1)
\}.
\end{eqnarray}
\begin{eqnarray}
\mathcal {F}_a^{LL,N}&=&-8 \pi  M^4 f_B C_f  r_{\psi }
\int_0^1dx_2dx_3\int_0^{\infty}b_2b_3db_2db_3\phi _D(x_3,b_3)\psi ^T(x_2,b_2)\nonumber\\&&
\{[ \left(r_c+\left(x_3-2\right) r_D\right)]E_{ef}(t_e)h(\alpha_a,\beta_e,b_2,b_3)\nonumber\\&&
+\left(x_2+1\right)r_DE_{ef}(t_f)h(\alpha_a,\beta_f,b_3,b_2)
\}.
\end{eqnarray}
\begin{eqnarray}
\mathcal {F}_a^{LL,T}&=&-8 \pi  M^4 f_B C_f  r_{\psi }
\int_0^1dx_2dx_3\int_0^{\infty}b_2b_3db_2db_3\phi _D(x_3,b_3)\psi ^T(x_2,b_2)\nonumber\\&&
\{[ \left(r_c+x_3
   r_D\right)]E_{ef}(t_e)h(\alpha_a,\beta_e,b_2,b_3)\nonumber\\&&
-\left(x_2-1\right)r_DE_{ef}(t_f)h(\alpha_a,\beta_f,b_3,b_2)
\}.
\end{eqnarray}
\begin{eqnarray}
\mathcal {M}_a^{LL,N}&=&\frac{8}{3} \pi  M^4  f_B C_fr_{J/\psi }
\int_0^1dx_2dx_3\int_0^{\infty}b_1b_2db_1db_2\phi _D(x_3,b_2)\exp(-\omega_B^2\frac{b_1^2}{2})\nonumber\\&&
\{[x_2r_{J/\psi }\psi ^V(x_2,b_2)]E_{gh}(t_g)h(\beta_g,\alpha_a,b_1,b_2)
-[  2 r_b r_D \psi
   ^T(x_2,b_2)\nonumber\\&&+\left(x_2-1\right) r_{J/\psi } \psi ^V(x_2,b_2)]E_{gh}(t_h)h(\beta_h,\alpha_a,b_1,b_2)
\}.
\end{eqnarray}
\begin{eqnarray}
\mathcal {M}_a^{LL,T}&=&\frac{8}{3} \pi  M^4 f_B C_f r_{J/\psi }^2
\int_0^1dx_2dx_3\int_0^{\infty}b_1b_2db_1db_2\phi_D(x_3,b_2)\psi ^V(x_2,b_2) \exp(-\frac{\omega_B^2 b_1^2}{2})\nonumber\\&&
\{x_2 E_{gh}(t_g)h(\beta_g,\alpha_a,b_1,b_2)
-(x_2-1)E_{gh}(t_h)h(\beta_h,\alpha_a,b_1,b_2)
\}.
\end{eqnarray}
\begin{eqnarray}
\mathcal {F}^{LL,N}_s&=&-2 \sqrt{\frac{2}{3}} \pi  M^4 f_B C_f f_{J/\psi } r_{J/\psi }
\int_0^1dx_2\int_0^{\infty}b_1b_2db_1db_2\phi _D (x_2,b_2)\exp(-\frac{\omega_B^2 b_1^2}{2})\nonumber\\&&\{
[  \left(r_b \left(4 r_D-1\right)-\left(x_2+1\right) r_D+2\right)]E_{ab}(t_{as})h(\alpha_{es},\beta_{as},b_1,b_2)S_t(x_2)\nonumber\\&&
+r_D
E_{ab}(t_{bs})h(\alpha_{es},\beta_{bs},b_2,b_1)S_t(x_1)]\},
 \end{eqnarray}
\begin{eqnarray}
\mathcal {F}^{LL,T}_s&=&-2 \sqrt{\frac{2}{3}} \pi  M^4 f_B C_f  f_{J/\psi } r_{J/\psi }
\int_0^1dx_2\int_0^{\infty}b_1b_2db_1db_2\phi _D (x_2,b_2)\exp(-\frac{\omega_B^2 b_1^2}{2})\nonumber\\&&\{
[\left(r_b-\left(x_2-1\right) r_D-2\right)]E_{ab}(t_{as})h(\alpha_{es},\beta_{as},b_1,b_2)S_t(x_2)\nonumber\\&&
-r_D
E_{ab}(t_{bs})h(\alpha_{es},\beta_{bs},b_2,b_1)S_t(x_1)]\},
 \end{eqnarray}
\begin{eqnarray}
\mathcal {M}_s^{LL,N}&=&\frac{8}{3} \pi  M^4 f_B C_f  r_{J/\psi }
\int_0^1dx_2dx_3\int_0^{\infty}b_1b_3db_1db_3\phi _D(x_2,b_1)\psi ^T(x_3,b_3)\nonumber\\&&\exp(-\omega_B^2\frac{b_1^2}{2})
\{(x_3-x_1)E_{cd}(t_{cs})h(\beta_{cs},\alpha_{es},b_3,b_1)\nonumber\\&&
+ \left(2
   \left(x_2+x_3-2\right) r_D-x_3-x_1+1\right)E_{cd}(t_{ds})h(\beta_{ds},\alpha_{es},b_3,b_1)
\},
\end{eqnarray}
\begin{eqnarray}
\mathcal {M}_s^{LL,T}&=&-\frac{8}{3} \pi  M^4 x_3 f_B C_f  r_{J/\psi }
\int_0^1dx_2dx_3\int_0^{\infty}b_1b_3db_1db_3\phi _D(x_2,b_1)\psi ^T(x_3,b_3)\nonumber\\&&\exp(-\omega_B^2\frac{b_1^2}{2})
\{(x_3-x_1)E_{cd}(t_{cs})h(\beta_{cs},\alpha_{es},b_3,b_1)\nonumber\\&&
-(x_1+x_3-1)E_{cd}(t_{ds})h(\beta_{ds},\alpha_{es},b_3,b_1)
\},
\end{eqnarray}
\begin{eqnarray}
\mathcal {F}^{LR,N}_s&=&-2 \sqrt{\frac{2}{3}} \pi  M^4 f_B C_f f_{J/\psi } r_{J/\psi }
\int_0^1dx_2\int_0^{\infty}b_1b_2db_1db_2\phi _D (x_2,b_2)\exp(-\frac{\omega_B^2 b_1^2}{2})\nonumber\\&&\{
[ \left(r_b \left(4 r_D-1\right)-\left(x_2+1\right) r_D+2\right)]E_{ab}(t_{as})h(\alpha_{es},\beta_{as},b_1,b_2)S_t(x_2)]\nonumber\\&&
-r_DE_{ab}(t_{bs})h(\alpha_{es},\beta_{bs},b_2,b_1)S_t(x_1)]\},
 \end{eqnarray}
\begin{eqnarray}
\mathcal {F}^{LR,T}_s&=&-2 \sqrt{\frac{2}{3}} \pi  M^4 f_B C_f  f_{J/\psi } r_{J/\psi }
\int_0^1dx_2\int_0^{\infty}b_1b_2db_1db_2\phi _D (x_2,b_2)\exp(-\frac{\omega_B^2 b_1^2}{2})\nonumber\\&&\{
[ \left(r_b-\left(x_2-1\right) r_D-2\right)]E_{ab}(t_{as})h(\alpha_{es},\beta_{as},b_1,b_2)S_t(x_2)]\nonumber\\&&
+r_DE_{ab}(t_{bs})h(\alpha_{es},\beta_{bs},b_2,b_1)S_t(x_1)]\},
 \end{eqnarray}
\begin{eqnarray}
\mathcal {M}_f^{LR,N}&=&\mathcal {M}_f^{LR,T}=\frac{8}{3} \pi  M^4 f_B C_f  r_{J/\psi }
\int_0^1dx_2dx_3\int_0^{\infty}b_1b_3db_1db_3\phi _D(x_3,b_3)\exp(-\omega_B^2\frac{b_1^2}{2})\nonumber\\&&
\{[ \left(\psi ^T(x_2,b_1)
   \left(r_c+x_2-1\right)-\left(x_2-1\right) r_{J/\psi } \psi ^V(x_2,b_1)\right)]\nonumber\\&&(E_{cd}(t_c)h(\beta_c,\alpha_e,b_3,b_1)
+E_{cd}(t_d)h(\beta_d,\alpha_e,b_3,b_1))
\}.
\end{eqnarray}
 \begin{eqnarray}
\mathcal {M}_a^{LR,N}&=&\mathcal {M}_a^{LR,T}=-\frac{8}{3} \pi  M^4 f_B C_f
\int_0^1dx_2dx_3\int_0^{\infty}b_1b_2db_1db_2\phi _D(x_3,b_2)\exp(-\omega_B^2\frac{b_1^2}{2})\nonumber\\&&
\{[ \left(x_3-1\right) r_D
   \psi ^V(x_2,b_2)-r_{J/\psi } \psi ^T(x_2,b_2) \left(2 r_c-x_2\right)]E_{gh}(t_g)h(\beta_g,\alpha_a,b_1,b_2)\nonumber\\&&
-[r_{J/\psi } \psi ^T(x_2,b_2)
   \left(-r_b+r_c+x_2-1\right)+r_D \psi ^V(x_2,b_2)
   \left(r_b+x_3\right)]\nonumber\\&&E_{gh}(t_h)h(\beta_h,\alpha_a,b_1,b_2)
\}.
\end{eqnarray}
\begin{eqnarray}
\mathcal {M}_s^{SP,N}&=&\frac{4}{3} \pi  M^4 f_B C_f r_{J/\psi }
\int_0^1dx_2dx_3\int_0^{\infty}b_1b_3db_1db_3\phi _D(x_2,b_1)\psi ^T(x_3,b_3)\exp(-\omega_B^2\frac{b_1^2}{2})\nonumber\\&&
\{[\left(2
   \left(x_2-x_3-1\right) r_D+x_3-x_1\right)]\nonumber\\&&E_{cd}(t_{cs})h(\beta_{cs},\alpha_{es},b_3,b_1)
-(x_1+x_3-1)E_{cd}(t_{ds})h(\beta_{ds},\alpha_{es},b_3,b_1)
\}.
\end{eqnarray}
\begin{eqnarray}
\mathcal {M}_s^{SP,T}&=&-\frac{4}{3} \pi  M^4 f_B C_f r_{J/\psi }
\int_0^1dx_2dx_3\int_0^{\infty}b_1b_3db_1db_3\phi _D(x_2,b_1)\psi ^T(x_3,b_3)\exp(-\omega_B^2\frac{b_1^2}{2})\nonumber\\&&
\{(x_3-x_1)E_{cd}(t_{cs})h(\beta_{cs},\alpha_{es},b_3,b_1)
-(x_1+x_3-1)E_{cd}(t_{ds})h(\beta_{ds},\alpha_{es},b_3,b_1)
\}.
\end{eqnarray}
\begin{eqnarray}
\mathcal {F}_a^{SP,N}&=&\mathcal {F}_a^{SP,T}=16\pi  M^4 f_B C_f
\int_0^1dx_2dx_3\int_0^{\infty}b_2b_3db_2db_3\phi _D(x_3,b_3)\nonumber\\&&
\{r_{J/\psi } \psi ^T(x_2,b_2)E_{ef}(t_e)h(\alpha_a,\beta_e,b_2,b_3)
+ r_D  \psi ^V(x_2,b_2)E_{ef}(t_f)h(\alpha_a,\beta_f,b_3,b_2)
\},\nonumber\\
\end{eqnarray}
where the expressions of  $\beta_{a,b,c,d}$ and $\alpha_e$ are the similar to those of Eq. (\ref{eq:betai11}),
 but with the replacement $r_{\eta_c}\rightarrow r_{J/\psi}$.

\section{scales and related functions in hard kernel }\label{sec:b}
We show here the functions $h$, coming from the Fourier transform  of virtual quark and gluon
propagators:
\begin{eqnarray}
h(\alpha,\beta,b_1,b_2)&=&h_1(\alpha,b_1)\times h_2(\beta,b_1,b_2),\nonumber\\
h_1(\alpha,b_1)&=&\left\{\begin{array}{ll}
K_0(\sqrt{\alpha}b_1) & \quad  \quad \alpha >0\\
K_0(i\sqrt{-\alpha}b_1)& \quad  \quad \alpha<0
\end{array} \right.\nonumber\\
h_2(\beta,b_1,b_2)&=&\left\{\begin{array}{ll}
\theta(b_1-b_2)I_0(\sqrt{\beta}b_2)K_0(\sqrt{\beta}b_1)+(b_1\leftrightarrow b_2) & \quad   \beta >0\\
\theta(b_1-b_2)J_0(\sqrt{-\beta}b_2)K_0(i\sqrt{-\beta}b_1)+(b_1\leftrightarrow b_2)& \quad   \beta<0
\end{array} \right.
\end{eqnarray}
where $J_0$ is the Bessel function and $K_0$, $I_0$ are modified Bessel function with
$K_0(ix)=\frac{\pi}{2}(-N_0(x)+i J_0(x))$.
The hard scale t is chosen as the maximum of the virtuality of the internal momentum transition in the hard amplitudes,
including $1/b_i(i=1,2,3)$:
\begin{eqnarray}
t_{a(as)}&=&\max(\sqrt{|\alpha_{e(es)}|},\sqrt{|\beta_{a(as)}|},1/b_1,1/b_2),\quad t_{b(bs)}=\max(\sqrt{|\alpha_{e(es)}|},\sqrt{|\beta_{b(bs)}|},1/b_1,1/b_2),\nonumber\\
t_{c(cs)}&=&\max(\sqrt{|\alpha_{e(es)}|},\sqrt{|\beta_{c(cs)}|},1/b_1,1/b_3),\quad t_{d(ds)}=\max(\sqrt{|\alpha_{e(es)}|},\sqrt{|\beta_{d(ds)}|},1/b_1,1/b_3),\nonumber\\
t_e&=&\max(\sqrt{|\alpha_a|},\sqrt{|\beta_e|},1/b_2,1/b_3),\quad t_f=\max(\sqrt{|\alpha_a|},\sqrt{|\beta_f|},1/b_2,1/b_3),\nonumber\\
t_g&=&\max(\sqrt{|\alpha_a|},\sqrt{|\beta_g|},1/b_1,1/b_2),\quad t_h=\max(\sqrt{|\alpha_a|},\sqrt{|\beta_h|},1/b_1,1/b_2).
\end{eqnarray}

The function $E_{ij}(t)$  is defined by
\begin{eqnarray}
E_{ab,cd,ef,gh}(t)=\alpha_s(t)S_{ab,cd,ef,gh}(t),
\end{eqnarray}
where the Sudakov factors can be written as
\begin{eqnarray}
S_{ab}(t)&=&s(\frac{M_B}{\sqrt{2}}x_1,b_1)+s(\frac{M_B}{\sqrt{2}}x_2,b_2)+s(\frac{M_B}{\sqrt{2}}(1-x_2),b_2)\nonumber\\&&
+\frac{5}{3}\int_{1/b_1}^t\frac{d\mu}{\mu}\gamma_q(\mu)+2\int_{1/b_2}^t\frac{d\mu}{\mu}\gamma_q(\mu),\nonumber\\
S_{cd}(t)&=&s(\frac{M_B}{\sqrt{2}}x_1,b_1)+s(\frac{M_B}{\sqrt{2}}x_2,b_1)+s(\frac{M_B}{\sqrt{2}}(1-x_2),b_1)\nonumber\\&&
+s(\frac{M_B}{\sqrt{2}}x_3,b_3)+s(\frac{M_B}{\sqrt{2}}(1-x_3),b_3)
\nonumber\\&&+\frac{11}{3}\int_{1/b_1}^t\frac{d\mu}{\mu}\gamma_q(\mu)+2\int_{1/b_3}^t\frac{d\mu}{\mu}\gamma_q(\mu),\nonumber\\
S_{ef}(t)&=&s(\frac{M_B}{\sqrt{2}}x_2,b_2)+s(\frac{M_B}{\sqrt{2}}(1-x_2),b_2)
+s(\frac{M_B}{\sqrt{2}}x_3,b_3)\nonumber\\&&+s(\frac{M_B}{\sqrt{2}}(1-x_3),b_3)
+2\int_{1/b_2}^t\frac{d\mu}{\mu}\gamma_q(\mu)+2\int_{1/b_3}^t\frac{d\mu}{\mu}\gamma_q(\mu),\nonumber\\
S_{gh}(t)&=&s(\frac{M_B}{\sqrt{2}}x_1,b_1)+s(\frac{M_B}{\sqrt{2}}x_2,b_2)+s(\frac{M_B}{\sqrt{2}}(1-x_2),b_2)
+s(\frac{M_B}{\sqrt{2}}x_3,b_2)\nonumber\\&&+s(\frac{M_B}{\sqrt{2}}(1-x_3),b_2)
+\frac{5}{3}\int_{1/b_1}^t\frac{d\mu}{\mu}\gamma_q(\mu)+4\int_{1/b_2}^t\frac{d\mu}{\mu}\gamma_q(\mu),
\end{eqnarray}
where the functions $s(Q,b)$ are defined in Appendix A of \cite{epjc45711}. $\gamma_q=-\alpha_s/\pi$ is the anomalous dimension of the quark.

\end{appendix}

\end{document}